\documentclass[fleqn, usenatbib]{mnras}
\usepackage[T1]{fontenc}
\usepackage{ae, aecompl}
\usepackage[dvipsnames]{xcolor}

\usepackage{graphicx}	
\usepackage{amsmath}	
\usepackage{amssymb}	

\newcommand{\kms}{\ifmmode {\rm km\,s}^{-1} \else km\,s$^{-1}$\fi} 
\newcommand{\ergs}{\ifmmode {\rm erg\,s}^{-1} \else erg\,s$^{-1}$\fi}

\newcommand\Msun{\ifmmode M_{\sun} \else $M_{\sun}$\fi}
\newcommand\msun{\ifmmode M_{\sun} \else $M_{\sun}$\fi}
\newcommand\Lsun{\ifmmode L_{\sun} \else $L_{\sun}$\fi}

\newcommand\mpyr{\ifmmode \Msun\,{\rm yr}^{-1} \else $\Msun\,{\rm yr}^{-1}$ \fi}  

\newcommand{\Ledd}{\ifmmode L_{\rm Edd} \else $L_{\rm Edd}$\fi}
\newcommand{\Luv}{\ifmmode L_{1350} \else $L_{1350}$\fi}
\newcommand{\lledd}{\ifmmode L/L_{\rm Edd} \else $L/L_{\rm Edd}$\fi}
\newcommand{\lamLlam}{\ifmmode \lambda L_{\lambda} \else $\lambda L_{\lambda}$\fi}
\newcommand{\Lbol}{\ifmmode L_{\rm Bol} \else $L_{\rm Bol}$\fi}

\newcommand{\mbh}{\ifmmode M_{\rm BH} \else $M_{\rm BH}$\fi}

\newcommand{\lem}{{ l_{\rm em} }} 
\newcommand{\zem}{{z_{\rm  em} }}

\newcommand{\vmath}[1]{\mathsf{v}_{\rm #1}} 
 
\newcommand{\vmathSym}[1]{\mathsf{v}_{#1}}

\defcitealias{Ch+H13}{Paper~I}
\defcitealias{MC97}{MC97} 
\defcitealias{MC98}{MC98} 
\defcitealias{BP82}{BP82} 
\defcitealias{EBS92}{EBS92} 
\defcitealias{VP06}{VP06}

\title[MHD Disc Winds and Line Width Distributions.II]{Magnetohydrodynamic Disc 
Winds and Line Width Distributions.II}
\author[L. S. Chajet and P. B. Hall]{L. S. Chajet$^{1}$ \thanks{E-mail: lchajet@yorku.ca}  
and P. B. Hall$^{1}$ \\
$^{1}$Department of Physics and Astronomy,
York University, Toronto, Ontario M3J 1P3, Canada} 

\date{Accepted XXX. Received YYY; in original form ZZZ}

\pubyear{2015}

\begin{document}
\label{firstpage}
\pagerange{\pageref{firstpage}--\pageref{lastpage}}
\maketitle

\begin{abstract}

We study AGN emission line profiles combining an improved version of the 
accretion disc-wind model of Murray \& Chiang  with the magneto-hydrodynamic 
model of Emmering et al. (1992).  
Here we extend our previous work to consider central objects with different 
masses and/or luminosities. 
We have compared the dispersions in our model  \ion{C}{iv} linewidth 
distributions to observational upper limit on that dispersion, considering 
both smooth and clumpy torus models. 
Following Fine et al., we transform that scatter in the profile line-widths into a 
constraint on the torus geometry and show how the half-opening angle of the 
obscuring structure depends on the mass of the central object and the accretion 
rate.  

We find that the results depend only mildly on the dimensionless angular momentum, 
one of the two integrals of motion that characterise the dynamics of the self-similar 
ideal MHD outflows. 
\end{abstract}


\section{Introduction}
Broad emission lines (BELs) are a spectral signature of Type 1 Active Galactic Nuclei (AGNs). 
Lines due to high-ionized species are generally single-peaked and blueshifted with 
res\-pect to the systemic velocity \citep[e.g.,][]{Sulentic+95, Sulentic+00, VandenBerk+01}.   
Although photoionization is well determined as the primary physical 
mechanism for their production, a detailed description of the region where 
the BELs originate is still an open question \citep[e.g.,][]{Chelouche+Zucker13, 
Grier+13, Sluse+11}. 
As the BLR is spatially unresolved, its structure and dynamics remain unclear and it is only 
through indirect methods that it can be stu\-died. 

Reverberation mapping (RM) is one of the few avai\-lable such methods. It relies on the 
time lag in the response of the BLR to variability of the continuum (see, e.g., the review 
by Peterson 2006).  
In particular, RM studies in seve\-ral AGNs have shown that the BLR is a stratified structure 
\citep[e.g.,][]{Peterson+Wandel00}:  within the same AGN, the high-ionization lines are 
produced closer to the central source than the low-ionization lines. 

Among the different competing and/or complementary models that have been developed 
to explain the nature of the BLR are discrete clouds and disc winds \citep[see 
brief reviews by e.g.,][]{Eracleous06, Everett07}. 
Broad-line cloud models describe this region as composed of numerous 
optically-thick clouds that, photoionized by the continuum-source emission, are the emitting 
entities responsible for the observed lines. 
This model successfully explains many of the observed spectral features, but also leaves 
several unsolved issues, such as the formation and confinement of the clouds.  

In models of outflowing gas, two major acce\-lerating mechanisms have been invoked: 
magnetohydrodynamic (MHD) driving \citep[e.g.,][]{BP82, EBS92, CL94, Kazanas+12}, 
and radiative acceleration \citep[e.g.,][]{MCGV95, Kurosawa+Proga09}. 
In some of these wind models the flow is assumed to be continuous 
\citep[e.g.,][]{KK94, MC97}, while in others it contains embedded inhomogeneities or 
clouds \citep[e.g.,][]{BKSB97}.  
Both of these types of models have had some success in fitting AGN observations. 
The spectral line shapes can be used as a tool to study the kinematics and 
physics of plasmas in AGNs.  
One additional important motivation of wind BLR models is that they provide a feasible 
feedback mecha\-nism to the host galaxy, thus regulating the co-evolution suggested  
by the solid relationships observatio\-nally found between the mass of the central object 
and some of the galaxy properties \citep[e.g.,][]{Marconi+Hunt03, Haring+Rix04, Gultekin+09}. 
It is also worth mentioning that the wind scenario offers the explanation least conflic\-ting
with the existence of a small group of AGN that show double-peaked line profiles
(generally, Balmer lines) in their spectra and with the fact that some among them  
fluctuate between a double- and a single-peaked profile 
\citep[e.g.,][]{Flohic+12, Eracleous+Halpern03}. 
In addition, the model high-velocity component of the wind naturally explains the
existence of blueshifted broad absorption lines seen in an optically selected subset 
of the quasar population, known as broad absorption line (BAL) quasars \citep{MCGV95, Elvis00}.

Because broad emission lines are produced close to the central engine, they carry 
important information on that region of the phenomenon \citep{Sulentic+00}.   
Hence, accu\-rate emission line models are important for studying AGN properties, 
such as black hole masses and accretion rates and put constraints on the unification 
paradigm. 
Indeed, due to their ubiquity, some authors \citep[e.g.,][]{Richards12} regard emission 
lines as a more powerful tool to studying AGN than absorption lines, because the latter 
are only present in a fraction of the quasar population. 

We study AGN emission line profiles combining an improved version of the accretion 
disc wind model of \citet{MC97} with the magnetohydrodynamic (MHD) dri\-ving of 
\citealt{EBS92}. 
Here we present the extension of our previous work in \citet[Paper I hereafter]{Ch+H13},  
considering a range of masses and luminosities.  
We have compared the averages in our model \ion{C}{iv} linewidth distributions to the 
observational results of \citet{Fine+10}, considering a smooth torus model. 
Additionally, we performed a similar study considering a clumpy torus \citep{N+08a}.  

The plan of the paper is as follows. In Section~\ref{sect: Model Description} we review our 
modifications to the \citet{MC97} disc-wind model.  
In Subsection~\ref{sect: MHD wind} we outline the basics of MHD winds, with emphasis on 
the description of \mbox{\citet{EBS92}} and show how we constructed our combined model. 
Section~\ref{sect: Line Profiles} presents the \ion{C}{iv} line profiles obtained with this model, 
and discusses their characteristics.  
In Section~\ref{sect: FWHM statistics}, using observational constraints on the dispersion 
of the line-widths, we derive constraints on the putative torus applicable to the various 
combinations of mass and Eddington ratio that we have considered. 
We explore both homogeneous (\ref{sect: smooth torus}) and clumpy torus (\ref{sect: clumpy torus}) 
prescriptions as the obscuring structure. 
In section \ref{sect: lambda 30}  we discuss how different values of the dimensionless angular 
momentum parameter $\lambda$ affect the results. 
We have considered two different values of $\lambda$ and run simulations using 
$\lambda = 10$ and $\lambda = 30$. 
These va\-lues were chosen to match those used by \citet{BP82} and \citet{EBS92}, respectively. 
In Section \ref{sect: discussion} we discuss our results and outline some possible ways 
for future exploration or improvement and in Section~\ref{sect: Conclusions} we present our 
conclusions.   
 
\section{Model Description} 
 \label{sect: Model Description} 
In \citetalias{Ch+H13} we presented our hybrid model, that combines the disc-wind model 
of \citet[hereafter MC97]{MC97}  with the MHD driving of \citet{EBS92} and here we 
include a brief description of it, that is applied to generate emission lines from AGNs. 

We assume an azimuthally symmetric, time-independent disc and wind. 
The lines arise at the base $z_{\rm em}$ of an emission region of thickness 
$\lem \ll \zem$, at an angle $\beta$ of the disc plane.  
This angle can, in principle, be a function of the radius, but we have adopted a 
fixed value.  
The wind streamlines make an angle of $\vartheta(r,z)$ relative to the disc plane. 
 
\begin{center}
\begin{figure} %
\centering
\includegraphics[width=\columnwidth]{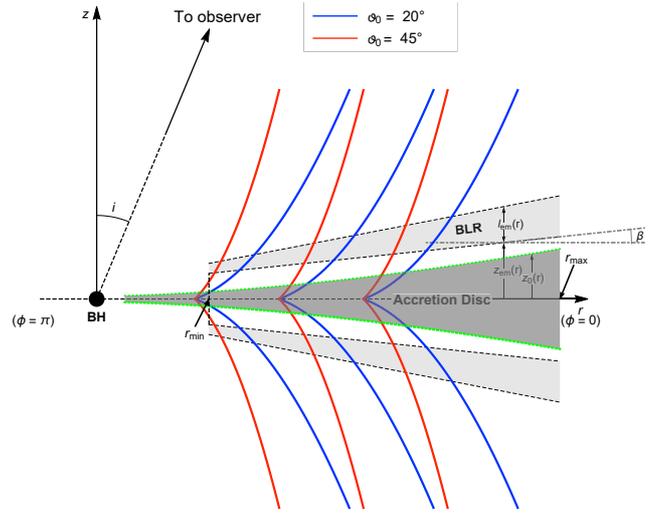} 
\caption[Streamlines for two different launch angles: $\vartheta_0=20^\degr, 45^\degr$]
{Streamlines for two different launch angles: $\vartheta_0=20^{\degr}, 45^{\degr}$. 
The disc midplane is assumed to lie in the $z=0$ plane and the disc is assumed to be 
opaque, such that the observer, at an angle $i$ from the $+z$ axis, sees only the 
upper quadrants. 
The lower dashed line represents the base of the emitting region, tilted by an angle 
$\beta$ with respect to the disc plane. 
The BLR is a layer of thickness $\lem(r)$ beginning a distance $\zem(r)$ 
above the disc midplane and spanning radii $r_{\rm min}< r < r_{\rm max}$.  
The height of the disc continuum photosphere is $z_0(r)$, where $r$ is the cylindrical 
radial coordinate. The disc wind lies at $z(r)>z_{\rm em}(r)$, and the emitting region 
for a given line is near the base of the wind, at 
$z_{\rm em}(r) < z < z_{\rm em}(r)+l_{\rm em}(r)$ above the disc midplane.}
\label{Fig: Different vartheta0 streamlines}
\end{figure}
\end{center}

Figure \ref{Fig: Different vartheta0 streamlines} depicts the system geometry and 
shows streamlines for two different launch angles. 

The line luminosity is given  by 
\begin{equation} 
\begin{split}
\label{eq: Lnu}
L_\nu (i) & = \int_{r_{\rm  {min}}}^{r_{\rm  {max}}} S_\nu(r)~\frac{r}{\cos\beta(r)} ~dr 
\int_0^{2\pi} \cos\iota(r,\phi,i)  ~ \times \\
&  (1-\exp[-\tau(r,\phi,i) \times e_\nu(r,\phi,i) \times e^{-x_\nu^2(r,\phi,i)}]) ~d\phi 
\end{split}
\end{equation}
where 
\begin{equation} 
\tau(r,\phi,i)  \equiv  
\frac{ck_0(r)/2\nu_0}{\sqrt{Q^2(r,\phi,i)+q_{\rm  {tt}}^2(r,\phi,i)}},   
\label{eq: tau} 
\end{equation}
\begin{equation}
e_\nu(r,\phi,i) \equiv
{\rm  {\, erfc}}
\left( 
-\frac{\nu-\nu_{\rm  {D}}(r,\phi,i)}
{\sqrt{2}\Delta \nu_{\rm  {tt}} \sqrt{1+q_{\rm  {tt}}^2(r,\phi,i)/Q^2(r,\phi,i)}} 
\right), \\
\label{eq: enu} 
\end{equation}
\begin{equation}
x_\nu^2(r,\phi,i) \equiv \frac{1}{2} \left(  
\frac{\nu-\nu_{\rm  {D}}(r,\phi,i)}{
\Delta \nu_{\rm  {tt}} \sqrt{1+Q^2(r,\phi,i)/q^2_{\rm  {tt}}(r,\phi,i)}}
\right)^2 
\label{eq: x2}  
\end{equation} 
and erfc is the complementary error function. 
The local inclination angle $\iota (i, r)$, that accounts for the different effective 
inclination that different portions of the emitting region have with respect to the 
line of sight (LoS), is defined by $\cos \iota = \cos i \cos \beta - \cos \phi \sin i \sin \beta$.  
The factor $k_0(r) \propto n(r)$ in expression \eqref{eq: tau} is the integrated line opacity 
(in units of Hz/cm), where $n(r)$ is the 
hydrogen number density evaluated at $z_{\rm  {em}}$. For the 
radial dependence of the density, we adopted $n(r)=n(r_{\rm min}/r)^2$ 
\citep[][MC98 hereafter]{MC98}.  
The term $Q$ in Equations \eqref{eq: tau} to \eqref{eq: x2} is a quadratic form that 
determines the projection of the gradient of the wind velocity along the LoS 
\citep[e.g.,][]{RH78, RH83}, according to 
\begin{equation} \label{eq: Q} 
Q \equiv \mathbf{\hat{n} \cdot \Lambda \cdot \hat{n}}, 
\end{equation}
where $\bf\hat{n}$ is the unit vector in the LoS direction and 
$\boldsymbol{\Lambda}$ is the strain tensor. 
The Doppler-shifted central frequency of the line emitted towards the observer 
from location ($r,\phi$) on the emitting surface is defined by 
$\nu_{\rm  {D}} = \nu_0 (1+ 
\vmath{D}/c)$, 
where $c$ is the speed of light in a vacuum. 

Expressions \eqref{eq: tau} to \eqref{eq: x2} include also the term `thermal $Q$', 
a quantity defined as the ratio of the characteristic thermal ($ \vmath{th}$) plus 
turbulent velocity ($ \vmath{turb}$) of the ion to the thickness of the emitting layer 
along the LoS:  
$q_{\rm  tt}(r, \phi,i)=   
\vmath{tt} \cos\iota(r,\phi,i)/l_{\rm  em}(r)\cos\beta$, with 
$\vmath{tt}^2 \equiv \vmath{th}^2 + \vmath{turb}^2 $. The turbulence is assumed to 
be isotropic. 
The effective frequency dispersion of the line is evaluated according to  
$\Delta \nu_{\rm tt} = \nu_0 \vmath{tt}/c$ and the emission 
region thickness is given by
\begin{equation} 
\label{eq: define_lem} 
\lem(r)=0.1{\zem}
\left[
\frac{\vmath{tt} + \vmath{p}(r, z_{\rm em})} 
{\vmath{tt}+\vmathSym{\infty}(r, z_{\rm em})}
\right],   
\end{equation} 
where $\vmath{p}(r,z)$ and $\vmathSym{\infty}(r, z)$
are the instantaneous and terminal poloidal velocities, respectively, of the wind along 
the streamline at $(r,z)$. 
The $z$-dependent quantities are evalua\-ted at $\zem$.  

The dependences of the quantities $q_{\rm  tt}(r, \phi, i)$ and 
$Q(r, \phi, i)$ on the different parameters are rather complex to make a clear 
comparison in the general case. We will, then, discuss this point in Section 
\ref{sect: Line Profiles}. 

\subsection{Ideal MHD wind launching}
 \label{sect: MHD wind} 
Magnetic acceleration of outflows has been often suggested in the literature as 
an efficient means of removing excess of angular momentum from the accreting 
material \citep[e.g.,][]{BP82, CL94, KK94, Everett05}.  
The standard non-relativistic ideal magneto-hydrodynamic (MHD) wind equations
are presented in (e.g.) the above mentioned references, as well as in 
\citetalias{Ch+H13}. 
In steady state  (i.e.  $\partial/\partial t = 0$) and assuming axisymmetry 
(i.e. $\partial/\partial \phi = 0$), the solutions have a number of conserved 
quantities along each magnetic field line \citep[e.g.,][]{Mestel68}, that 
impose cons\-traints to the flow dynamics. These are  the mass to magnetic 
flux ratio, $\frac{k}{4 \pi} = \frac{\rho \; \vmath{p}}{B_{\rm  p}}$;  
specific angular momentum,  
$l = r \left(\vmath{\phi} - \frac{B_\phi}{k} \right)$ 
and specific energy, 
$e =\frac{\boldsymbol{\mathsf v}^2}{2} + h + \Phi_{\rm  g} -
 \frac{ r  \Omega B_{\phi}  }{k}$, where the subscripts p and $\phi$ correspond 
respectively to the poloidal and azimuthal components of the velocity, 
$\boldsymbol{\mathsf v}$, and magnetic, $\boldsymbol{B}$, 
vector fields; $\rho$ is the gas density; $\Phi_{\rm  g}$, the gravitational potential; 
and $h$ and $\Omega$ are the specific enthalpy and the angular velocity, 
respectively.

An important quantity defined for magnetized fluids is the so-called 
Alfv\'en Mach number.  
The square of this quantity is defined as $m =\vmath{p}^2/\vmath{Ap}^2$, where 
$\vmath{Ap}$ is the poloidal component of the Alfv\'en velocity  
$\boldsymbol{\mathsf v}_{\rm  A} = \frac{\boldsymbol{B}}{\sqrt{4 \pi \; \rho}}$
which is the characteristic velocity of the propagation of magnetic signals in 
an MHD fluid.  
The Alfv\'en radius $r_{\rm A}$ is the point on each poloidal field line where 
$m = 1$; the loci of all $r_{\rm A}$ define the Alfv\'en surface. 
The subscript ``A'' refers, hereafter, to quantities evaluated at the Alfv\'en point.  

\subsubsection{Blandford \& Payne MHD Wind Solution} 
\label{sect: BP82 model}
Analytic solutions for the system of equations that defines the stationary ideal 
MHD problem can be obtained only under some simplifying assumption such 
as self-similarity \citep[e.g.,][BP82 hereafter]{BP82}.  
This solution for the field can be written in terms of variables $\chi$, $\xi(\chi)$, 
$\phi$, and $r_0$,  which are related to the cylindrical coordinates via
\begin{equation}\label{eq: r}
\boldsymbol{r} \equiv \left[r, \phi, z\right] = \left[r_0 \xi(\chi), \phi, r_0 \chi \right],
\end{equation}
where the adopted independent variables $(r_0, \chi)$ are a pair of spatial coordinates 
analogous to $(r, z)$.
The function $\xi(\chi)$ describes the shape of the field lines and, in the general case, 
is not \textit{a priori} known, but found as part of a self-consistent solution to the MHD 
equations. 
The flow velocity components are given by
\begin{equation}\label{eq: v}
\boldsymbol{\mathsf v} = \left[\xi'(\chi)f(\chi), g(\chi), f(\chi) \right] \sqrt{\frac{G M}{r_0}},
\end{equation}
where a prime denotes differentiation with respect to $\chi$, and $G$ and $M$ 
are respectively the gravitational constant and the mass of the central black hole.  

The self-similar scaling of the magnetic field amplitude $B$, and gas density $\rho$ 
with the spherical radial coordinate $r$ is given, in the general case 
\citep[e.g.,][]{EBS92, CL94, Kazanas+12}, by $\rho  \propto r_0^{-b}$, for which 
$B \propto r_0^{-(b+1)/2}$. 
The \citetalias{BP82} solution is a particular case corresponding to $b = 3/2$. 
The magnetic field and density at arbitrary positions can be then written, in 
accordance with the self-similarity {\it Ansatz} (Equation \eqref{eq: r}), as 
${\boldsymbol{ B} }= B_0(r_0) {\boldsymbol{b}}(\chi)$ 
and $\rho = \rho_0 (r_0) \varrho(\chi)$.
On the disc plane the rotational velocity, $\vmathSym{\phi}$, is Keplerian and scales 
as $\vmathSym{\phi} \propto r_0^{-1/2}$. 
The functions $\xi(\chi)$, $f(\chi)$ and $g(\chi)$ have to satisfy the flow MHD equations 
subject to the above scalings of $\rho$, $B$, and $\vmathSym{\phi}$ and boundary conditions. 
In particular, at the disc surface $\xi(0)=1$, $f(0)=0$ and $g(0)=1$.

The parameters of the model are the dimensionless expressions of the integrals of motion: 
$\epsilon =e/ (GM/r_0)$, $\lambda = l/(GMr_0)^{1/2}$ and 
\mbox{$\kappa =  k(1+\xi'^2_0) \left[(GM/r_0)^{1/2}/B_{0}\right]$}. 
An extra parameter is the value of the derivative $\xi'_0$ of the self-similar function $\xi$ at 
the disc surface.   
However, due to the regularity conditions that must be satisfied, these parameters are not 
independent.
In particular, the value of $\xi'_0\equiv \xi'(\chi=0)$ must be chosen to ensure the regularity 
of the solution at the Alfv\'en point. 
That reduces the degrees of freedom of the solutions, that are then parametrized by 
two numbers. 

\citetalias{BP82} studied outflows that become super-Alfv\'enic (i.e.,  $m > 1$) at 
a finite height above the disc. The asymptotic conditions for such flows admit two 
kind of solutions that depend on the location of the fast-mode Mach number $n$ 
above the disc. 

\begin{figure} 
\includegraphics[width=\columnwidth]{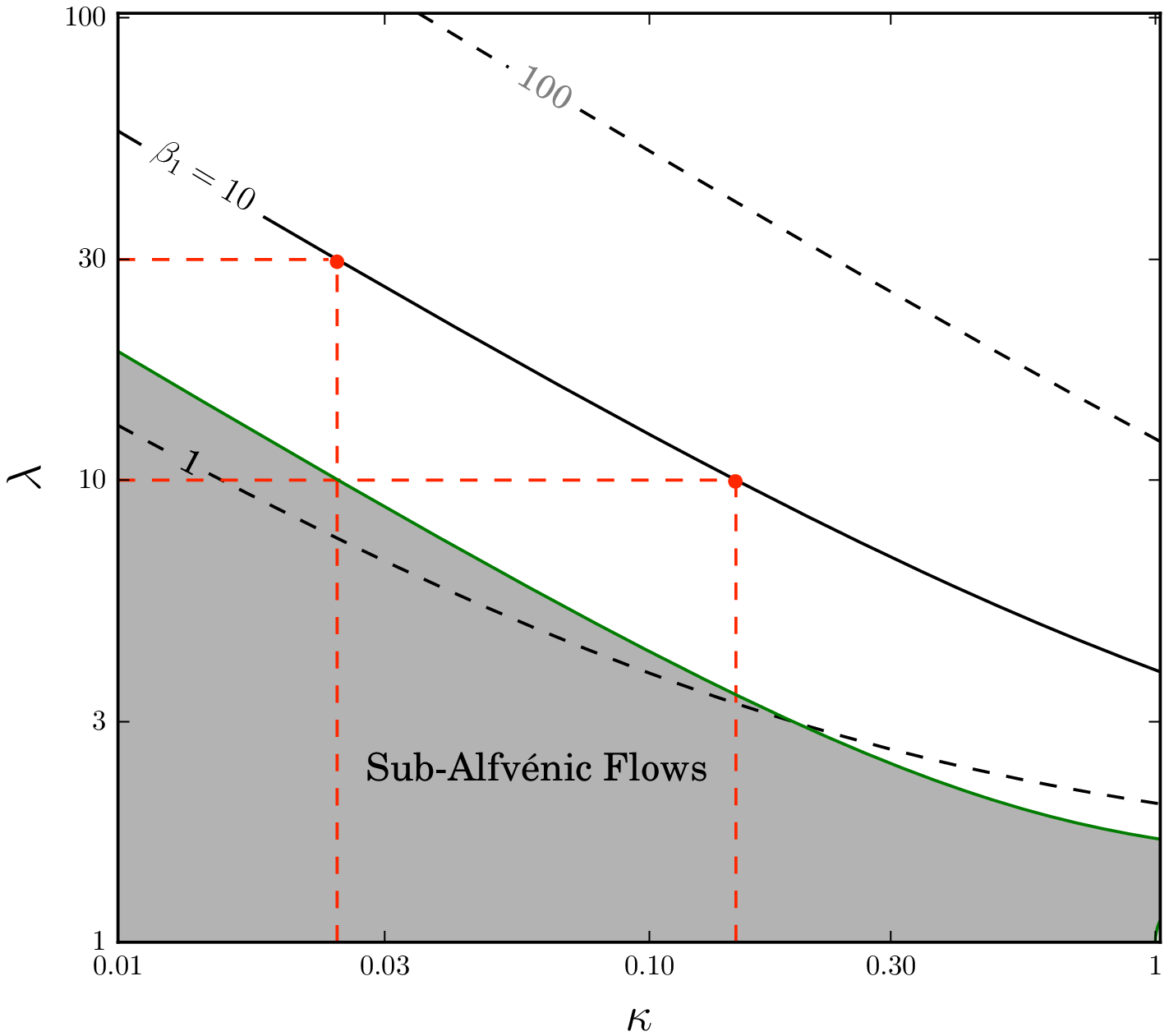} 
\caption{Contour plot analogous to Fig. 2 of \citet{BP82}. Each contour corresponds 
to a value of the quantity $\beta_1$, defined in Equation \eqref{eq: BP82eq2.31}, 
that is fixed for a given problem.  
The solid contour corresponds to $\beta_1 = 10$ %
and the possible ($\kappa, \lambda$) pairs for our problem, set by the 
condition in Equation \eqref{eq: kappa_lambda} imposed by the \citet{EBS92} 
choice of parameters. 
The dashed red lines locate the two points (also marked in red) for which we 
have explored in this work. 
Flows in the grey-shadowed region satisfy the condition  
$\kappa \lambda_{\rm min}(2 \lambda_{\rm min} - 3)^{1/2} < 1$, and will never 
become super-Alfv\'enic ($4 \pi \rho~\vmath{p}^2 > B_{\rm p}^2$ ).  
} 
\label{Fig:simBP82-Fig2}
\end{figure}

The family of solutions that asymptotically approach $n =1$ as $\chi \rightarrow \infty$  
have the generic form $\xi =  c_1 \chi^{\alpha}$, where $c_1$ and $\alpha = 1-3^{3/2}/\beta_1$ 
are constants, controlled by the parameter 
\begin{equation}
\label{eq: BP82eq2.31} 
\beta_1  \equiv \kappa \left(2 \lambda - 3\right)^{3/2} \gg 1.     
\end{equation}
Figure \ref{Fig:simBP82-Fig2} (similar to Figure 2 in BP82) depicts the contours of 
constant $\beta_1$ for three different values of the parameter. 

The asymptotic value of the function $f$ is given by 
\begin{equation} 
\label{eq: f_infty}
f_\infty =  \left(\frac{2 \lambda-3}{3}\right)^{1/2}.
\end{equation} 
%
\subsubsection{Emmering et al. model} 
 \label{subsect: self-sim sol.} 

\citet[][EBS92 hereafter]{EBS92} impose {\it a priori} the functional form of the solution 
so that it will asymptotically tend to the type of \citetalias{BP82} solution presented above.      
In their equation (3.19), \citetalias{EBS92} give an explicit form for the function $\xi(\chi)$:
\begin{equation} \label{eq:xi}
    \xi = \left( \frac{\chi}{c_2} +  1\right)^{1/2}, 
\end{equation} 
where $c_2 = \frac{1}{2}\tan \vartheta_0$ was chosen to ensure that the field lines
make an initial angle $\vartheta_0$ with the disc plane, so that
$\cot \vartheta_0 = \xi'_0$ and the subscript $0$ means that the quantities are 
evaluated at the disc plane.
It can be shown  \citep[e.g., BP82,][]{Heyvaerts96} that this angle 
must satisfy the condition $\vartheta_0 < 60^{\circ}$.

For the form proposed by \citetalias{EBS92}, the only possi\-ble value of $\beta_1$ 
is $6 \sqrt{3} \sim 10$. 
Replacing this value in Equation \eqref{eq: BP82eq2.31} gives that, as 
shown by \citetalias{EBS92}, for this kind of solution the parameters $\kappa$ and 
$\lambda$ must be related by: 
\begin{equation} \label{eq: kappa_lambda}
\kappa = 2 \left(\frac{3}{2 \lambda-3}\right)^{3/2}
\end{equation}
Thus, in this model the solutions depend on $\lambda$ and $\vartheta_0$. 

The \citetalias{EBS92} work was intended to support the cloud scenario of the BLR, 
and thus the authors  proposed that the emission lines arise in clouds confined by 
an MHD flow. 
However,  we follow \citetalias{MC97} and \citetalias{MC98} in assuming that the 
lines form in a continuous medium. 
We consider line emissivity obtained by a CLOUDY photoioni\-zation model, different 
from either of the emissivity laws adopted by \citetalias{EBS92}. Two of those emissivity 
models include electron scattering, which is not considered in our model.   
In \citetalias{EBS92} both the dimensionless angular momentum $\lambda$ and the 
launch angle are fixed. We, on the other hand, varied those parameters to study their 
effect on the profiles.   
Note also that, while \citetalias{EBS92} obtains the line luminosity integra\-ting in the 
two poloidal variables, we include the $z$-integral in the optical depth expression, 
already incorporated in Equation \eqref{eq: tau}.   

\section{Line Profiles} 
\label{sect: Line Profiles} 
In  \citetalias{Ch+H13} we applied the modified model described above to our fiducial 
case, characterised by mass $M = 10^8 \Msun$ and luminosity  $\Luv$ (defined below) 
$10^{46} \ergs$, and considered a dimensionless specific angular momentum 
$\lambda = 10$.  
We then studied several combinations of inclination angles, in the range 
$5\degr \leqslant i \leqslant 84\degr$ and launch angles 
$6\degr \leqslant \vartheta_0 < 60\degr$. The maximum viewing angle was set as to be 
the angle at which an observer would see the base of the emitting region, that was chosen 
to make an angle $\beta = 6^{\circ}$ from the disc plane. 
In the present work we analyse, for the same  set of viewing and launch angles, how 
the emission line profiles for different masses and luminosities  are effected.   
For each of such combinations, the specific luminosity from each component of the 
\ion{C}{iv} doublet is computed separately, and then the results added together.  

To scale the inner and outer radii of the emitting region from the fiducial case to any 
mass-luminosity combination we adopted a luminosity- and mass-based scaling:   
\begin{equation} 
\label{eq: def_rmin-rmax}
r_{\rm min, \;max} =  2 M_8^{1/3} L_{46}^{0.5} \; 10^{15, \,19} \;\, \text{cm}, 
\end{equation}
where $M_8 =  \frac{M}{10^8 \Msun}$ and 
$L_{46} = \frac{\Luv}{10^{46} \ergs}$ and $M$ is the mass, in units of 
solar masses and $\Luv$ is the object's UV luminosity, in \ergs. 
We had initially set a scaling depen\-ding only on luminosity, 
$r_{\rm min, \;max} =  2 L_{46}^{0.5} \; 10^{15, \,19}$ cm. Howe\-ver, the profiles rendered 
by such a scaling were too broad and  had much larger linewidth dispersions than the 
fiducial case.  
Incorporating the mass dependence to the scaling had the effect of reducing both the 
FWHMs and corres\-ponding dispersions. 
A  mass-dependent scaling has been applied before (in the optical by e.g., \citealt{Flohic+12}). 
In that work, the computational domain (inner and outer radii of the BLR) is given in units 
of the gravitational radius of the central object, $R_{\rm G} = G M/c^2$.  
The mass scaling adopted here is motivated by the results on the BLR reported by 
\citet{Elitzur+14}. 

As discussed in  \citetalias{Ch+H13}, we determined the source function for our simulations 
by applying the RM results of \citet{Kaspi+07} to the radial line luminosity function $L(r)$ 
calcula\-ted by \citetalias{MC98} for a quasar with 
$\Luv \equiv \nu L_\nu(1350 \;\! \text{\AA}) =10^{46}\, \ergs$ and shown in their Figu\-re 5b.  
The radii in the \citetalias{MC98} Figure 5b line luminosity function have been empirically 
adjusted down by a factor of five to match the RM results of \citet{Kaspi+07}.  
Other relevant physical parameters are the density and density power-law exponent and 
thermal plus turbulent velocity, chosen to be $n_0=10^{11}$ cm$^{-3}$, $b=2$ and 
$ \vmath{tt} = 10^7$ cm s$^{-1}$, respectively, same as the 
corresponding values used in  \citetalias{Ch+H13}.  

We are now in position to compare the quantities $q_{\rm  tt}(r, \phi, i)$ and $Q(r, \phi, i)$.  
AS mentioned above, a general description is not possible, but we can discuss the 
limiting cases that we have identified. Running our code for a number of combinations 
of relevant parameters shows that the behaviour of the ratio 
$R_{\rm Qq} = \left| \frac{Q(r, \phi, i)}{q_{\rm  tt}(r, \phi, i)} \right|$ depends strongly 
on $\vartheta_0$. For small values of $\vartheta_0$ ($ \lesssim 30\degr$), and given 
$i$, the ratio is a strong function of the radius, with some azimuthal modulation at any 
given radius. The azimuthal modulation level decreases with decreasing inclination. 
For the smallest inclinations ($i \lesssim 10\degr$), there is a turnover radius beyond 
which the overall dominant factor changes. 
For example, the $(\vartheta_0, i) = (6\degr, 10\degr)$ combination gives 
$1 \lesssim  R_{\rm Qq} \lesssim 53$ for most ($\sim 2/3$) of the radial domain and 
$0.17 \lesssim R_{\rm Qq} \lesssim 1$ for the largest radii. 

On the contrary, for large values of $\vartheta_0$ ($\gtrsim 30\degr$), the dominant 
factor in $R_{\rm Qq}$ is $q_{\rm  tt}$. As an example, for 
$(\vartheta_0, i) = (45\degr, 10\degr)$ $R_{\rm Qq}$ lies in the range 
$0.0031 \lesssim R_{\rm Qq} \lesssim 0.17$. 
Exact values of the ratio depend on the radius and increase with increasing $i$. 
The azimuthal modulation is only important at large inclination values ($i\geq45\degr)$. 
Although the values reported correspond to the fiducial case, the qualitative 
description is correct for the different masses and accretion rates studied. 
 
\begin{figure*} 
\includegraphics[width=\textwidth, clip=false]{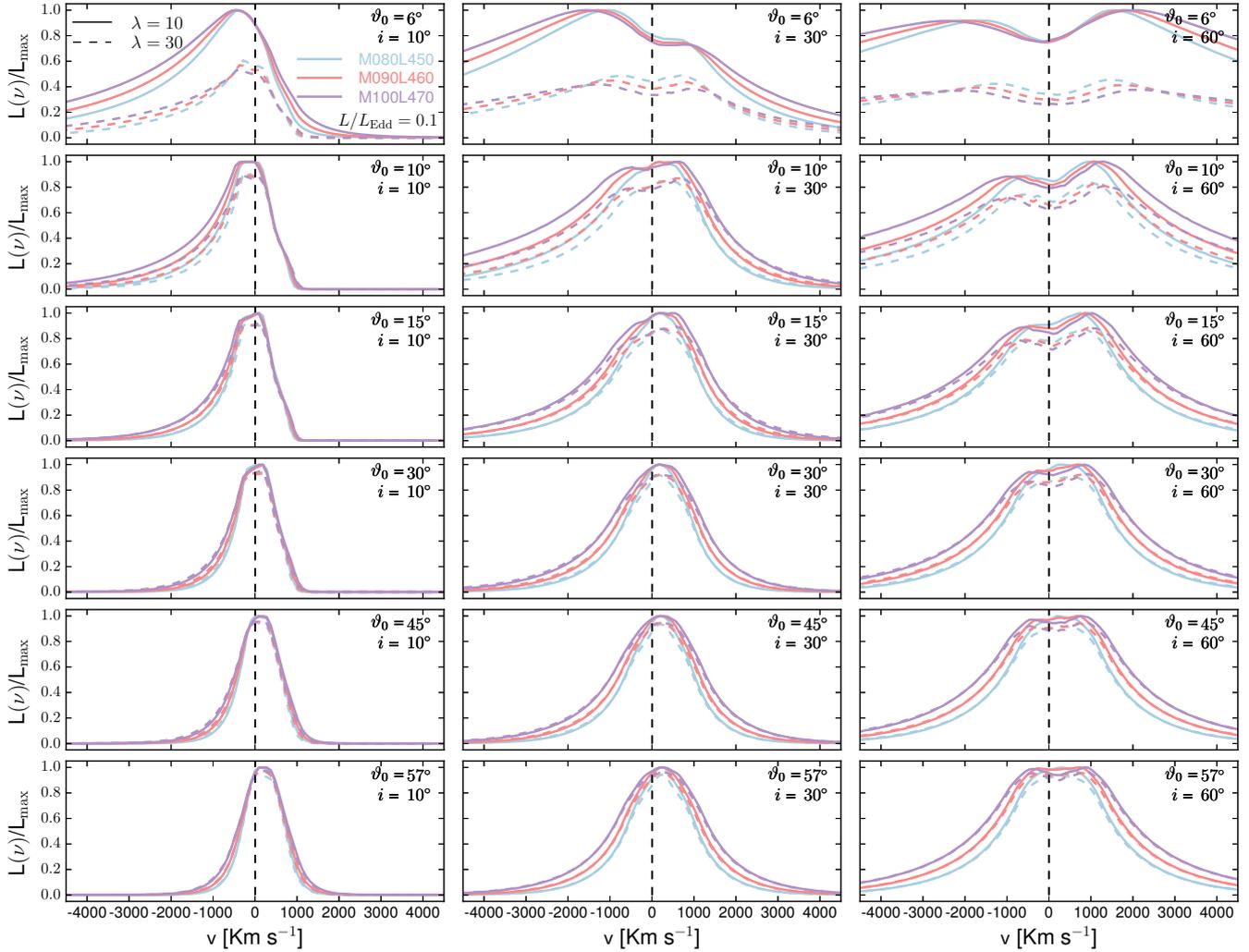} 
\caption{Normalized (with respect to the corresponding $\lambda = 10$ case profile) line 
luminosity vs. velocity for several values of viewing and launch angles. 
Each panel shows the profiles for given launch and viewing angles. Along rows, 
launch angles are constant, and viewing angles increase from left to right. 
Only profiles corresponding to three viewing angles ($i = 10^{\circ}, 30^{\circ}, 60^{\circ}$) 
are shown. 
The labels indicate $\log M_i$ and $\log L_i$, where $M_i$ is the mass of the central 
BH in the corresponding simulation, in units of solar masses,  and $L_i$ is the assumed 
luminosity in \ergs. For example, M080L450 indicates $\log M_i = 8.0$ and 
$\log L_i =  45.0$. 
This set of lines correspond to the case $\lledd = 0.1$. 
}
\label{Fig: LineProfiles-Lambda10-And-30-AccrRate01}
\end{figure*}

\begin{figure*} 
\includegraphics[width=\textwidth, clip=false]{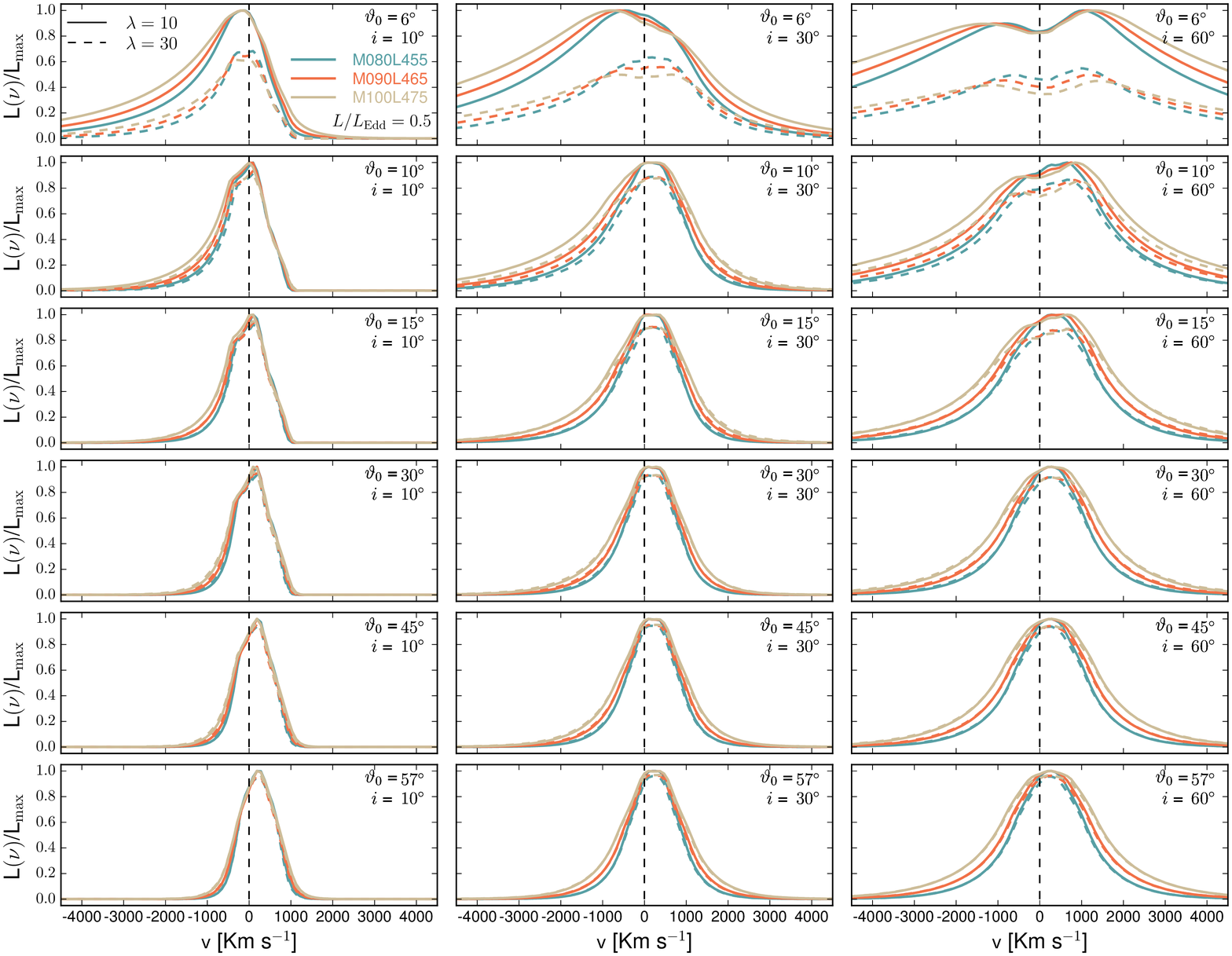} 
\caption{Normalized (with respect to the corresponding $\lambda = 10$ case profile) line 
luminosity vs. velocity for several values of viewing and launch angles. Similar to 
Fig. \ref{Fig: LineProfiles-Lambda10-And-30-AccrRate01}, but for $\lledd = 0.5$.
}
\label{Fig: LineProfiles-Lambda10-And-30-AccrRate05}
\end{figure*}

\begin{figure*} 
\includegraphics[width=\textwidth, clip=false]{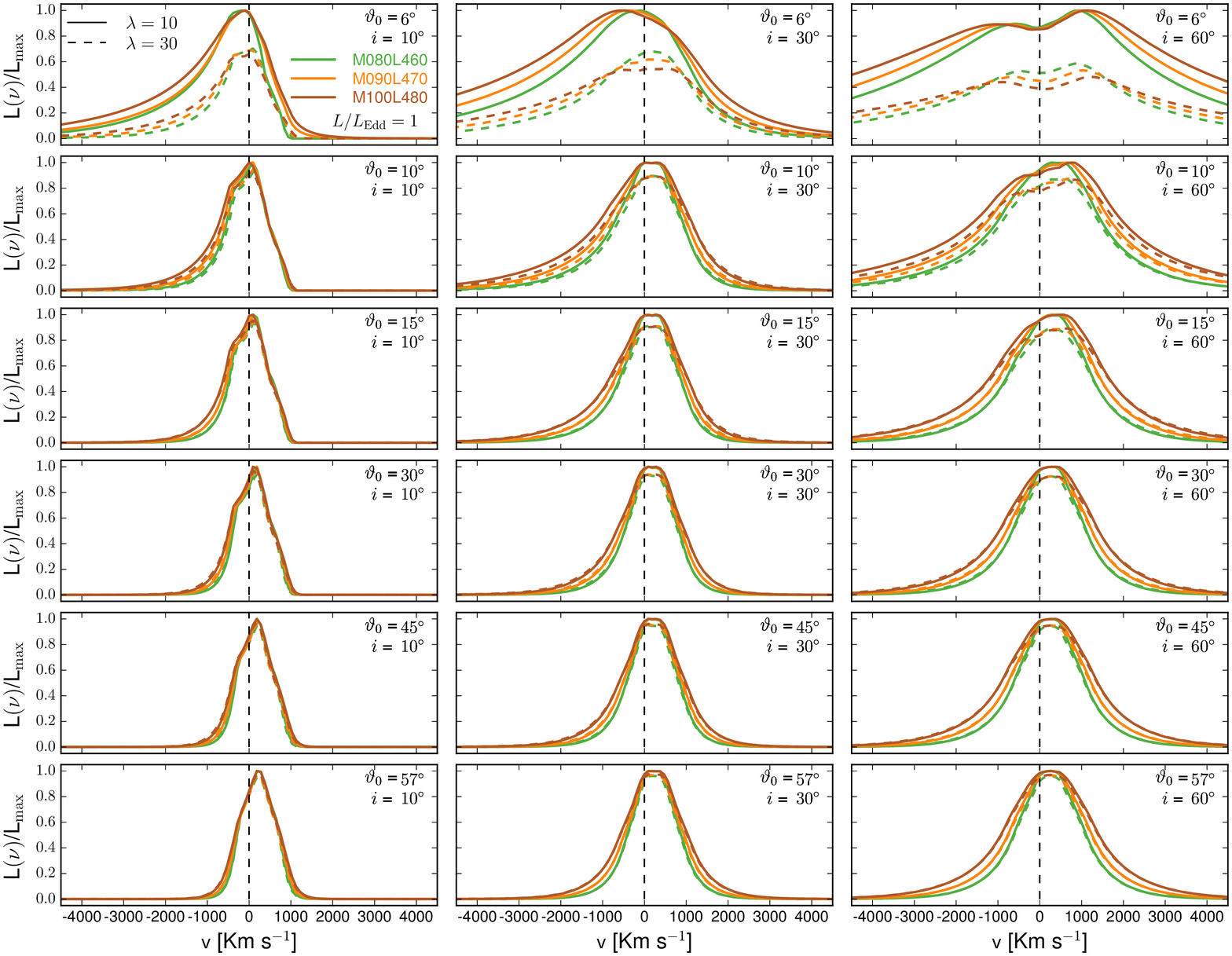} 
\caption{Normalized (with respect to the corresponding $\lambda = 10$ case profile) line 
luminosity vs. velocity for several values of viewing and launch angles. Similar to 
Fig. \ref{Fig: LineProfiles-Lambda10-And-30-AccrRate01}, but for $\lledd = 1$.
}
\label{Fig: LineProfiles-Lambda10-And-30-AccrRate1}
\end{figure*}

The line profiles are determined by the kinematics and geometry of the gas distribution in 
the emission region. The inner and outer radii of that region, defined 
by Equation (\ref{eq: def_rmin-rmax}), can be recast to show the dependence on the 
accretion rate. Figures \ref{Fig: LineProfiles-Lambda10-And-30-AccrRate01} to 
\ref{Fig: LineProfiles-Lambda10-And-30-AccrRate1} show several profiles obtained 
in our simu\-lations. The diffe\-rence from one figure to the other is the value of the 
accretion rate, as shown by the inside legend. 
The Eddington ratio, that measures the accretion rates in Eddington units, is defined as  
$\dot{m} = L_{\rm Bol}/L_{\rm Edd}$, where $L_{\rm Edd}$ is the Eddington luminosity 
given by 
$
L_{\rm Edd} = 1.26 \times 10^{46} \ergs$ ($M_{\rm  BH}/10^8\Msun)
$. 
 
Each panel of these figures includes the lines correspon\-ding to fixed viewing and 
launch angles and luminosities, as des\-cribed in the caption to 
Figure \ref{Fig: LineProfiles-Lambda10-And-30-AccrRate01}.
Solid and dashed lines correspond to $\lambda = 10$ and 30, respectively. The 
normalization is with respect to the corresponding $\lambda = 10$ case and only 
results correspon\-ding to a subset of the viewing angles in the simulations are included.  
We postpone until section \ref{sect: lambda 30} the discussion on how different 
$\lambda$ values affect the  results. 
The velocities in the $x$-axes represent velocities from the observer's point of view,
therefore negative velocities corres\-pond to blueshifts. The zero velocity, marked by 
a vertical dashed line in each panel, is the average 
of the two doublet wavelengths.  

Within each panel, not only the profiles are broader with increasing mass and 
decreasing luminosity, but also the shift of the peak and the fractions of the 
flux blue- and redward of the central velocity vary.   For fixed $\vartheta_0$, the 
profiles are broader as the inclination angle increases. 
The dependence of the line-profile width on mass and luminosity 
can be seen by considering the Keplerian speed $\vmath{K} \sim \sqrt{M/r}$      
and using Equation (\ref{eq: def_rmin-rmax}) to express $r \sim M^{1/3} L^{1/2}$, 
we find that  $\vmath{K} \sim M^{1/3} L^{-1/4}$.
The profiles corres\-ponding to any mass-luminosity combination have similar general 
behaviour and trends to the fiducial case, analysed in \citetalias{Ch+H13}.  
For instance, all profiles have a certain degree of asymmetry that decreases with 
increasing inclination. 
As before, the blue wings change less than the red wings, so that as the inclination 
angle increases, the red wings become relatively stronger.  
%
 
Physically the wind velocity depends on the launch angle, but the observer sees the 
fraction projected onto the LoS.  Howe\-ver, as discussed in \citetalias{Ch+H13}, 
in the framework of the present model, the actual angle to be considered is the angle 
$\vartheta$ at which a line launched with some $\vartheta_0$ crosses the base of the 
emitting region (when $\vartheta_0$ increases, so does $\vartheta$).  
For any given combination of the two angles $i$ and $\vartheta_0$, the projection of the 
velocity will be towards the observer on the fractions of the emitting region that satisfy the 
condition $\vartheta > i$. 
The velocity relevant for producing the observed line profiles is the Doppler velocity, 
that includes a contribution from the rotational velocity. 
As $\vartheta$ increases, the rotational velocity becomes increa\-singly dominant, producing 
more symmetric profiles.  
For any launch angle, the relative importance of the receding term with respect to the 
approaching term increases with increasing viewing angle, but the effect is larger for smaller 
$\vartheta_0$. 

The shape of a spectral line can be characterised  by analysing the ratio 
$S = \text{FWHM/}\sigma_{\rm  l}$, where $\sigma_{\rm  l}$ is the standard deviation of the line. 
For a Gaussian profile, $S_{\rm  {Gauss}} =  2 \sqrt{2 \ln (2)} \simeq 2.35$, while 
$S \rightarrow 0$ for Lorentzian and logarithmic profiles. 
For our results we find minimum and mean $S$ values $\sim 0.69$ and $\sim 1.33$ respectively. 
%

\begin{figure*} 
\includegraphics[width=\textwidth, clip=false]{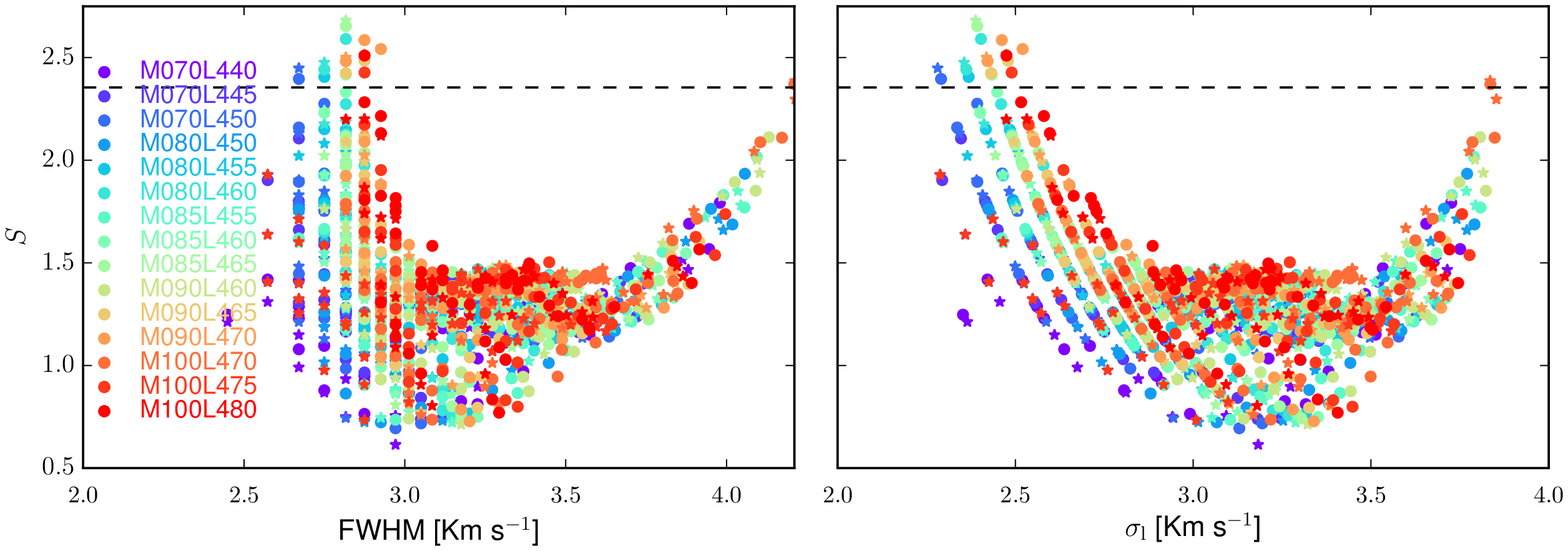}  
\caption{                
Line shape parameter, $S$ vs. FWHM (left) and $S$ vs. $\sigma_{\rm l}$ for all 
$\lambda =10$ profiles, with $S = \text{FWHM/}\sigma_{\rm  l}$, where 
$\sigma_{\rm  l}$ is the standard deviation of the line. 
The dashed line shown in both panels indicates 
$S = S_{\rm  {Gauss}} = 2 \sqrt{2 \ln (2)} \simeq 2.35$. Labels follow the scheme laid 
out in the caption to Figure \ref{Fig: LineProfiles-Lambda10-And-30-AccrRate01}. 
}
\label{Fig:LineRatios-vs-Incl-Lambda10}
\end{figure*}

%
Figure \ref{Fig:LineRatios-vs-Incl-Lambda10} shows the distribution of the parameter $S$ 
for our profiles as a function of two different line-width measures: FWHM in the right and 
$\sigma_{\rm  l}$ in the left.  
The fact that most of our lines are below the $S_{\rm  {Gauss}}$ value,  indicates that they 
have more prominent wings than a simple Gaussian profile. This is not surprising, as  
the dynamics is dominated by the turbulent plus thermal velocity.  

\section{FWHM statistics} 
\label{sect: FWHM statistics} 
Here we explore the FWHM of our profiles in relation to the luminosity of the sources and 
then consider the dispersion of this line-width characterisation. 
Using the FWHM allows a fairly simple comparison to the observational data and constraints 
reported by \citet{Fine+10}, who used interpercentile values (IPV) to charac\-terise the 
line-widths, and an immediate possibility of extension to 
other works in the literature. 

\begin{figure*} 
\centering 
\includegraphics[width=\textwidth]{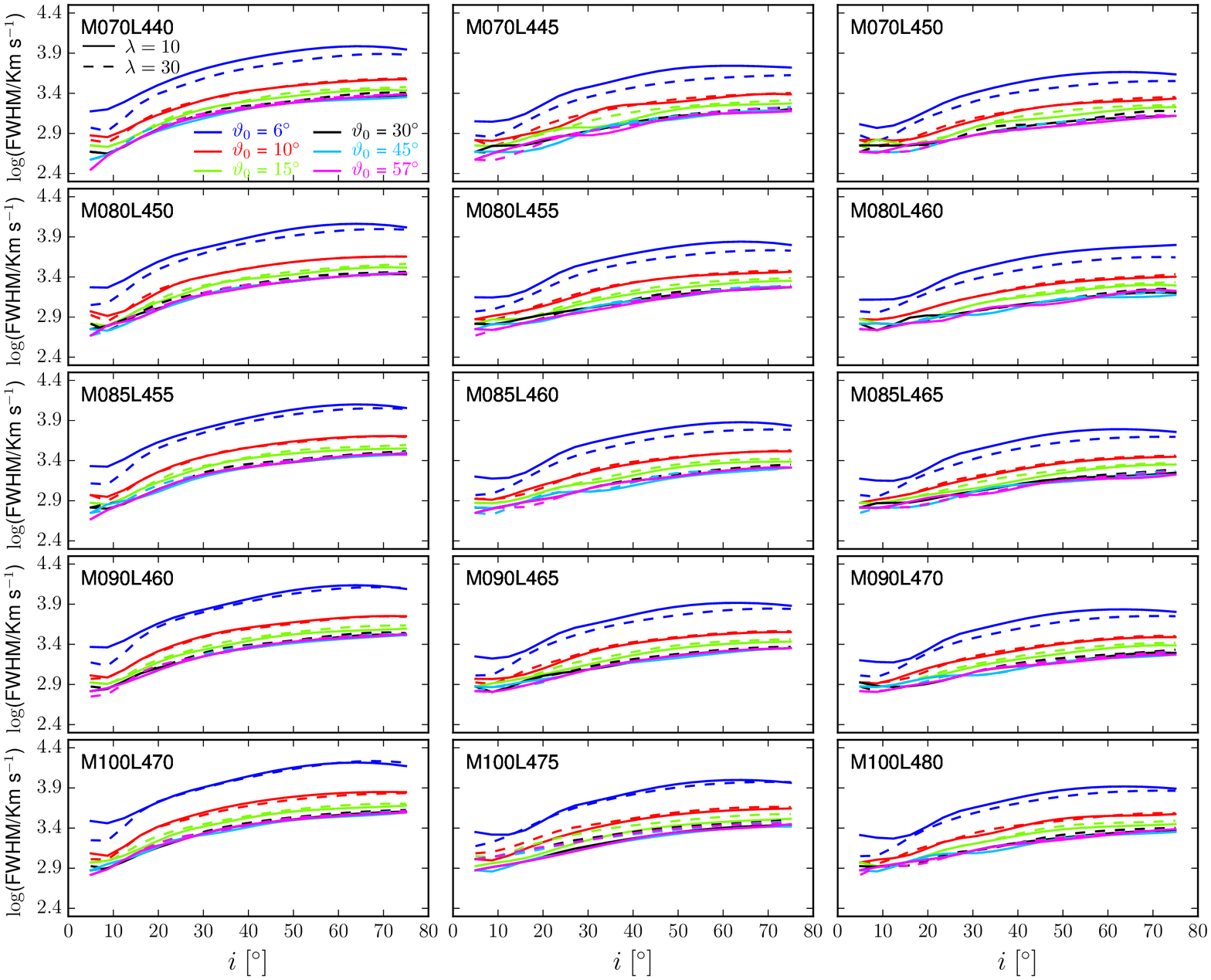} 
\caption{Profile FWHM vs. inclination angle. Solid lines are $\lambda = 10$ 
cases, while $\lambda = 30$ cases are represented by dashed lines. 
Each panel corresponds to a different combination of mass and luminosity 
and each line represents results for a given launch angle. 
} 
\label{Fig: Log10FWHMvsIncl} 
\end{figure*}

Figure \ref{Fig: Log10FWHMvsIncl} shows the $\log(\text{FWHM})$ of the line 
profiles from our simulations. Each panel corresponds to a given mass and luminosity  
and each colour, to a different launch angle, with solid and dashed lines used for 
the $\lambda = 10$ and $\lambda = 30$ cases respectively. 
Here we can see again that the results follow the expected behaviour of the 
FWHM with mass and luminosity, i.e., the FWHM increases with increasing mass 
and decreasing luminosity.   

Notably, for any given mass and luminosity (i.e., within a particular panel), the 
differences due to the angular momentum are almost independent of the 
launch angle, except for the lowest $\vartheta_0$ value.  As mentioned, 
detailed discussion on this matter is placed in section  \ref{sect: lambda 30}.  

\subsection{Luminosity-linewidth plane} 
\label{sect: L-FWHM plane}
\begin{figure*} 
\includegraphics[width=\textwidth]{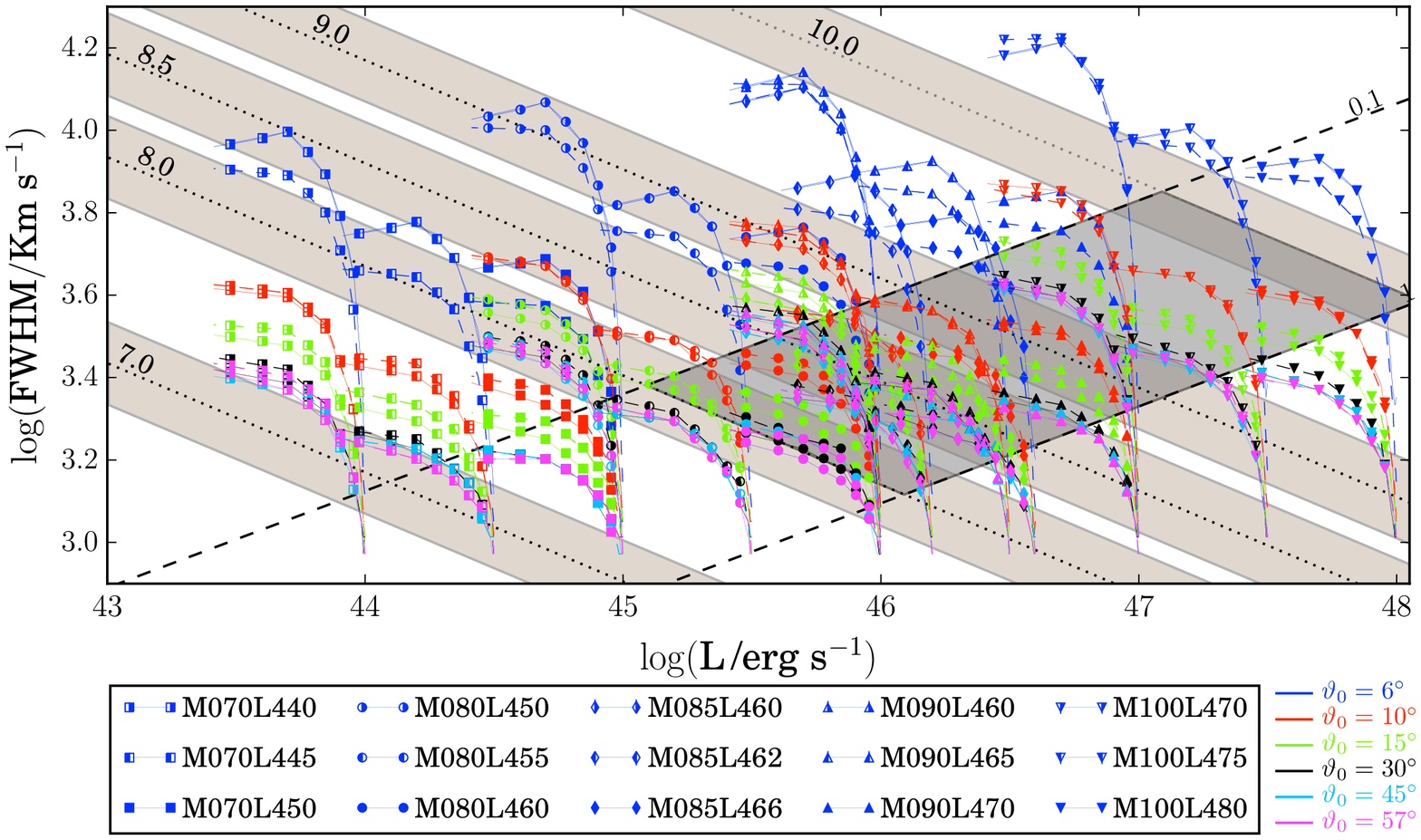} 
\caption{
Similar to Figure 2 of \citet{Fine+10}. The grey-shaded area corresponds approximately 
to the densest region in that figure.  
Note that the quantities in both axes differ from the ones used by \citet{Fine+10}. 
In the $x$-axis, we used luminosities instead of absolute magnitudes and 
for the line-width measure, in the $y$-axis, we used FWHM instead of IPV. 
The dotted and dashed lines represent loci of constant mass and Eddington ratio, respectively, 
where the mass has been evaluated using Equation (7) in \citetalias{VP06} 
(reproduced in Equation (\ref{eq: VP06-eq7})). 
The shaded regions are bound by lines of constant
$\log(M_i/M_{\sun}) \pm 0.2$, where $M_i$ is one of the mass values used in 
our simulations. 
Any point in the plane represents the result from a simulation with given mass, luminosity 
and launch and inclination angles. 
Each symbol corresponds to a mass-luminosity combination, 
and each colour (with the same scheme used in other figures), to a launch angle. 
For fixed $\vartheta_0$ the coloured solid lines join cases corresponding to different 
viewing angles. 
For any given mass-luminosity case, we see that results from the lowest $\vartheta_0$ 
values are consistently larger, for fixed $i$, that results from any other launch angle.   
Furthermore, the lower the launch angle, the larger the spread in $\log{(\mbox{FWHM})}$ 
values. 
}
\label{Fig: simFig2-Fine10}
\end{figure*}

Figure \ref{Fig: simFig2-Fine10} is similar to Figure 2 from \citet{Fine+10}. 
The grey-shaded area corresponds approximately to the densest region in that figure. 
We recall here  again that we used line FWHM (instead of IPV) and in the $x$-axis, 
luminosities have been plotted, instead of absolute magnitudes.  
The dotted and dashed lines represent loci of constant mass and Eddington ratio, 
respectively, where the mass has been evaluated, as \citet{Fine+10}, using Equation (7) 
in \citet[VP06, hereafter]{VP06}, reproduced below: 
\begin{align}
\label{eq: VP06-eq7} 
\log{\mbh (\ion{C}{iv})} = \log{  
\left\{
\left[ 
\frac{\text{FWHM(\ion{C}{iv})}}{1000 \kms}
\right]^2  
\left[ 
\frac{\lambda L_\lambda (\text{1350 \!\AA}) }{10^{44} \ergs} 
\right]^{0.53}  \nonumber 
\right\} 
} \\ 
&  \hspace{-6.75cm} + (6.66 \pm 0.01). 
\end{align}

The shaded regions are bound by lines of constant $\log(M_i) \pm 0.2$, where $M_i$ is 
one of the mass values used in our simulations, in units of solar masses.   
While the sample standard deviation of the weighted average zero point of their mass 
scaling relationships reported by \citetalias{VP06} is 0.36 dex, we used a smaller 
range around $\log M_i$ to avoid overlapping of regions belonging to two different 
such va\-lues.  
For each mass-luminosity combination, results corresponding to the same launch angle 
but different inclination are linked by solid lines, with the same colour scheme used in 
other figures.  

The differences in luminosity from sources of intrinsic identical luminosities are attributable 
to the differences in viewing angle, because the observer sees a fraction that depends on 
the projection on the sky of the emitting region, so the $x-$axis is constructed by multiplying 
the assumed object's luminosity by the cosine of the inclination angle.  

\begin{figure} 
\includegraphics[width=\columnwidth]{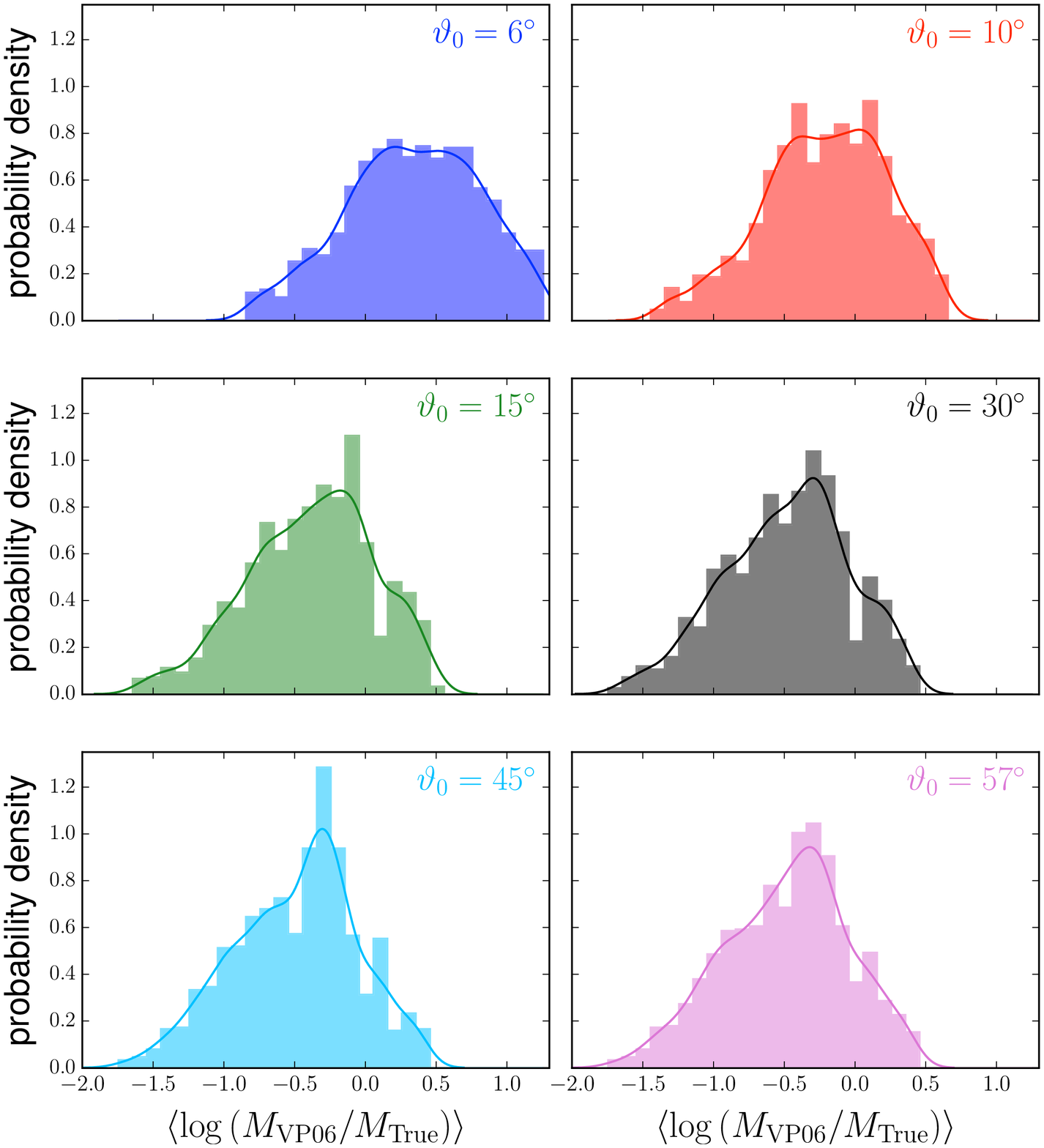} 
\caption{Normalized histograms of the quantity $\langle \log{(M_{\rm VP06}/M_{\rm True})} \rangle$,  
up to a set of $i_{\rm max}$ values within a smooth torus model, for the different mass sets. 
Each panel corresponds to a given launching angle.  
The solid lines represent smooth histograms (kernel density estimation, KDE) over the same data.  
The skewness is positive in all cases, except the corresponding to $\vartheta_0 = 6\degr$, and 
becomes increasingly negative as the launch angle increases. 
} 
\label{Fig: Data-Histogram-mass-ratio-6} 
\end{figure}

\begin{figure} 
\includegraphics[width=\columnwidth]{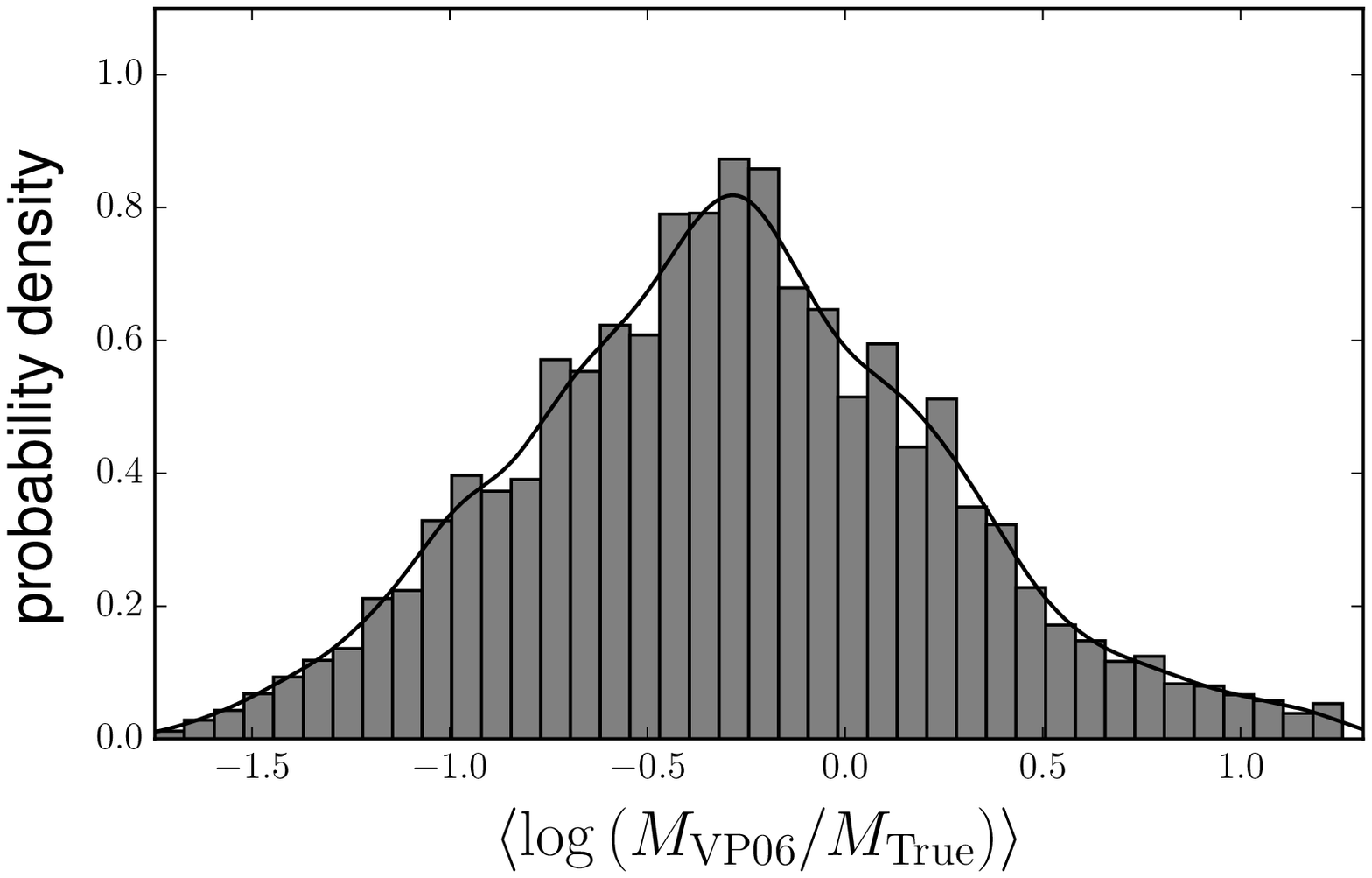} 
\caption{Normalized histograms of the quantity $\langle \log{(M_{\rm VP06}/M_{\rm True})}\rangle$ 
as a function of $i_{\rm max}$ for all mass sets, no discriminating by launch angle 
or black hole mass.   
} 
\label{Fig: Data-Histogram-mass-ratio}
\end{figure}

Note that if the black hole mass were estimated using Equation \eqref{eq: VP06-eq7}, many 
of the results could not retrieve their original mass values. In general, the cases that seem 
best represented by the expression are in the range $8.0 \lesssim \log M_i \lesssim 8.5$, 
in particular for the largest viewing angles.   
For $\log M_i \sim 9$, fewer combinations, corresponding to the larger Eddington ratios, 
smaller $\vartheta_0$ and larger $i$, reproduce the original mass values, whereas 
$\log M_i \sim 10$, no combination can retrieve their true \mbh. 
This is partly due to the fact that the expression given by \citetalias{VP06} does not include 
any angular dependency, although the issue is, in effect, analysed in their work.  
While the case of dependence on launch angle has not been studied, the influence of the 
viewing angle has been considered in several works 
\citep[e.g.,\citetalias{VP06};][]{Collin+06, Decarli+10, Assef+11, Denney12}. 
One way to compare the masses obtained by applying ,\citetalias{VP06} Equation 7 
versus the objects' true masses is to study, for given $\vartheta_0$, the mean of the ratio 
of the two quantities up to different $i_{\rm max}$ values in a torus model. 
Figure \ref{Fig: Data-Histogram-mass-ratio-6} shows normalized histograms of 
$\langle \log{(M_{\rm VP06}/M_{\rm True})} \rangle$, with the panels arranged according to the 
launch angle. The FWHM values are taken from the interpolated function. All histograms 
are skewed and their widths depend strongly on the launch angle.  
Note that the skewness direction also depends on this angle, and is positive for the smallest 
$\vartheta_0$ and becomes increasingly negative as the launch angle increases. 
In Figure \ref{Fig: Data-Histogram-mass-ratio} we present the normalized histogram of the distribution 
of $\langle \log{(M_{\rm VP06}/M_{\rm True})} \rangle$ as function of  $i_{\rm max}$ for the whole 
set adopting a smooth torus model (discussed in more detail in section \ref{sect: smooth torus}). 

Note that the correlation between FWHM and IPV has a non-negligible scatter, as 
can be seen from e.g., Figu\-re A8 of \citet{Fine+10}. Moreover, there is no direct 
conversion from one measure to the other, except for well-determined cases, such 
as a Gaussian curve, for which the relation is $\mbox{FWHM} = 1.75\, \mbox{IPV}$. 
This degene\-racy makes the comparison between our linewidth vs. luminosity with 
Figure 2 of \citet{Fine+10} not completely straightforward. 
A more direct comparison can be done with the results of \citet{Decarli+08}, who 
reported a mean $\langle \mbox{FWHM} \rangle = 4030 \pm 1200$ \kms from their 
sample. 
This value might be consistent with our findings, although it should be pointed 
out that their sample was much smaller than that of \citet{Fine+10}.    

\subsection{Dispersion of $\log{\text{FWHM}}$-smooth torus} 
\label{sect: smooth torus} 

\begin{figure} 
\centering
\begin{center} 
\includegraphics[width=\columnwidth]{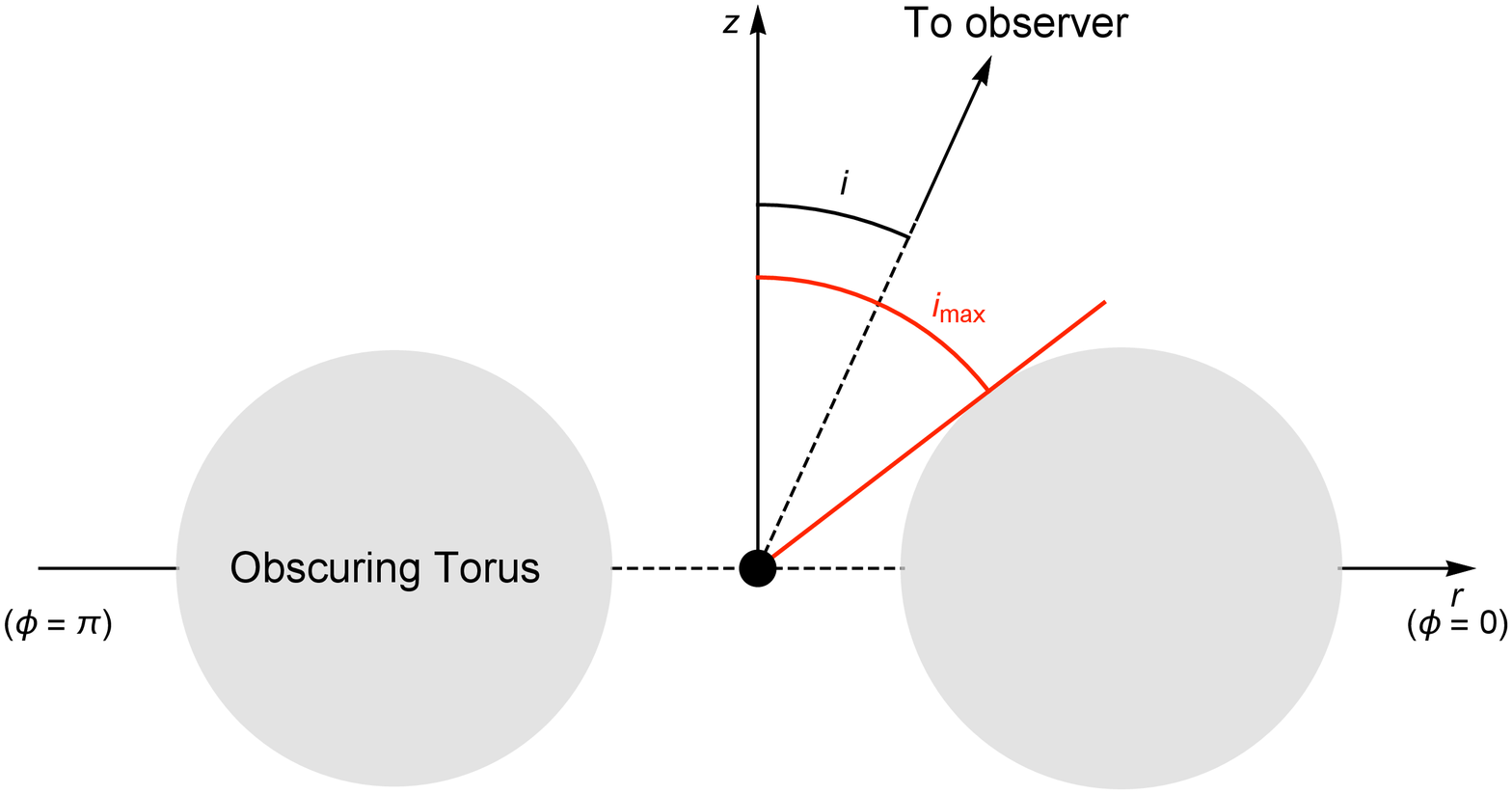} 
\caption{
Torus geometry. The angle $i$ is the observing angle to the AGN, and the opening 
angle $i_{\rm  max}$ assumed to be constrained by an obscuring torus. 
}
\label{Fig: Fine08-Fig13}
\end{center}
\end{figure}

Analogous to what we did in  \citetalias{Ch+H13}, we also analyse the FWHM of our line 
profiles applying the prescription of \cite{Fine+08, Fine+10}, who constrained the possible 
viewing angles using geometrical models for the BLR 
and comparing the expected dispersion in line-widths to their observational data. 
Figure \ref{Fig: Fine08-Fig13} sketches the assumed geometry.  

Using the launch angle as a parameter, we evaluate the dispersion of the function 
$f(i)=\log (FWHM)$. 
For a {\textbf {given $i_{\rm min}$}, the mean and the variance of the FWHMs are 
functions of $i_{\rm max}$, according to 
\begin{align}
\bar{f}(i_{\rm  {max}})  = &
\frac{ \int_{i_{\rm  {min}}}^{i_{\rm  {max}} }  \sin{i} \; P(i)\; f(i) \, di}
{ \int_{i_{\rm  {min}}}^{i_{\rm  {max}} }  \sin{i} \; P(i) \, di}  
\label{eq: mean-f} \,, \\[.7cm]
\sigma^2_f(i_{\rm  {max}})  = &   
\frac{ \int_{i_{\rm  {min}}}^{i_{\rm  {max}} }  \sin{i} \; P(i) \;  \left[ f(i)-\bar{f}(i_{\rm  {max}}) \right ]^2 \, di}
{ \int_{i_{\rm  {min}}}^{i_{\rm  {max}} }  \sin{i} \; P(i)\, di} \, ,  
\label{eq:disp-f}
\end{align} \\ 
where, for the case of a smooth obscuring torus, the escape probability is a step 
function 
\begin{equation*}
P(i) = \left\{ 
\begin{array}{l}
1 \qquad \text{for $i \leqslant i_{\rm  {max}}$} \\ 
0 \qquad  \mbox{for $i >  i_{\rm  {max}}$} 
\end{array}
\right.
\end{equation*}
for a given $ i_{\rm  {max}}$.  
 
Here we perform the same calculations for each of the combinations of mass and 
luminosity. 
Thus, for each set of profiles obtained for different inclination angles $i$ and launch 
angles, $\vartheta_0$ we study how the FWHMs are distributed with respect to $i$. 
For the \ion{C}{iv} case that is of interest here, \cite{Fine+10}~obtained an observational 
$\sigma_f(i_{\rm max}) = 0.08$ dex limit.  

We show the dispersion of our line profile sets versus $i_{\rm max}$ in Figure 
\ref{Fig: VarvsInclAngle}. The panels are arranged as constant mass along 
rows and constant Eddington ratio along columns, and the black horizontal 
dashed line in each of them represents the \cite{Fine+10} constraint. 
Only dispersions that sa\-tisfy $\sigma_f(i_{\rm max}) \leqslant 0.08$ dex are 
allowed, which imposes a cons\-traint on the possible $i_{\rm max}$. 
In this test, the constraint on $i_{\rm max}$ represents the torus half-opening 
angle that would yield a Type 1 object.  
The solid and dashed lines correspond to $\lambda = 10$ and $\lambda = 30$ 
cases, respectively. 

The set of allowed $i_{\rm max}$ changes from left to right, increasing as the 
Eddington ratio increases. 
From top to \mbox{bottom}, i.e., for constant Eddington ratio, the range of allo\-wed 
$i_{\rm max}$ values shrinks with increasing mass. 
These features are related to the torus geometry and its dependence on mass 
and luminosity and will be discussed below. 
\begin{figure*} 
\centering
\includegraphics[width=\textwidth]{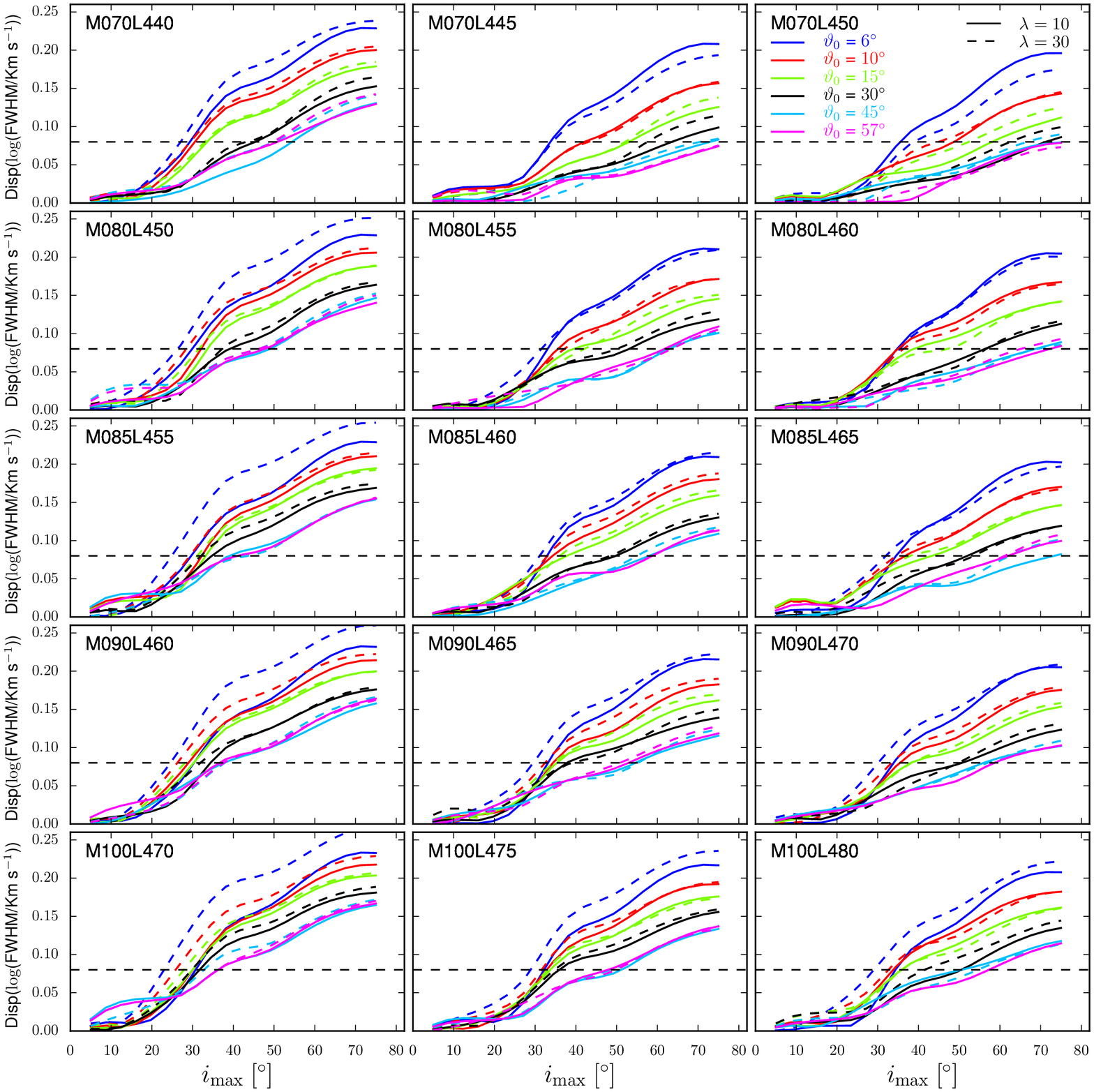} 
\caption{Standard deviation of the profile FWHM vs. inclination angle. 
Each panel corresponds to a different combination of mass and 
luminosity and each line represent results for a given launch angle. 
The black horizontal dashed line corresponds to the \citet{Fine+10} 0.08 dex 
constraint. 
} 
\label{Fig: VarvsInclAngle}
\end{figure*} 

As shown in  \citetalias{Ch+H13} for the fiducial case, another way to visua\-lise 
the constraining of the $i_{\rm max}$ parameter is by ma\-king, for each 
mass-luminosity combination, contour plots of the corresponding 
$\sigma_f(i_{\rm max})$ as function of both $\vartheta_0$ and $i_{\rm max}$. 
In Figure \ref{Fig: ContourPlot-i_maxVsTheta0-Lambda10} we show such contour 
plots of the standard deviation of the profile FWHM vs. launch and inclination 
angles for our set of masses and luminosities for the case $\lambda = 10$.   
The data are shown in the same arrangement as in Figure \ref{Fig: VarvsInclAngle}.
 
\begin{figure*} 
\centering 
\includegraphics[width=\textwidth]{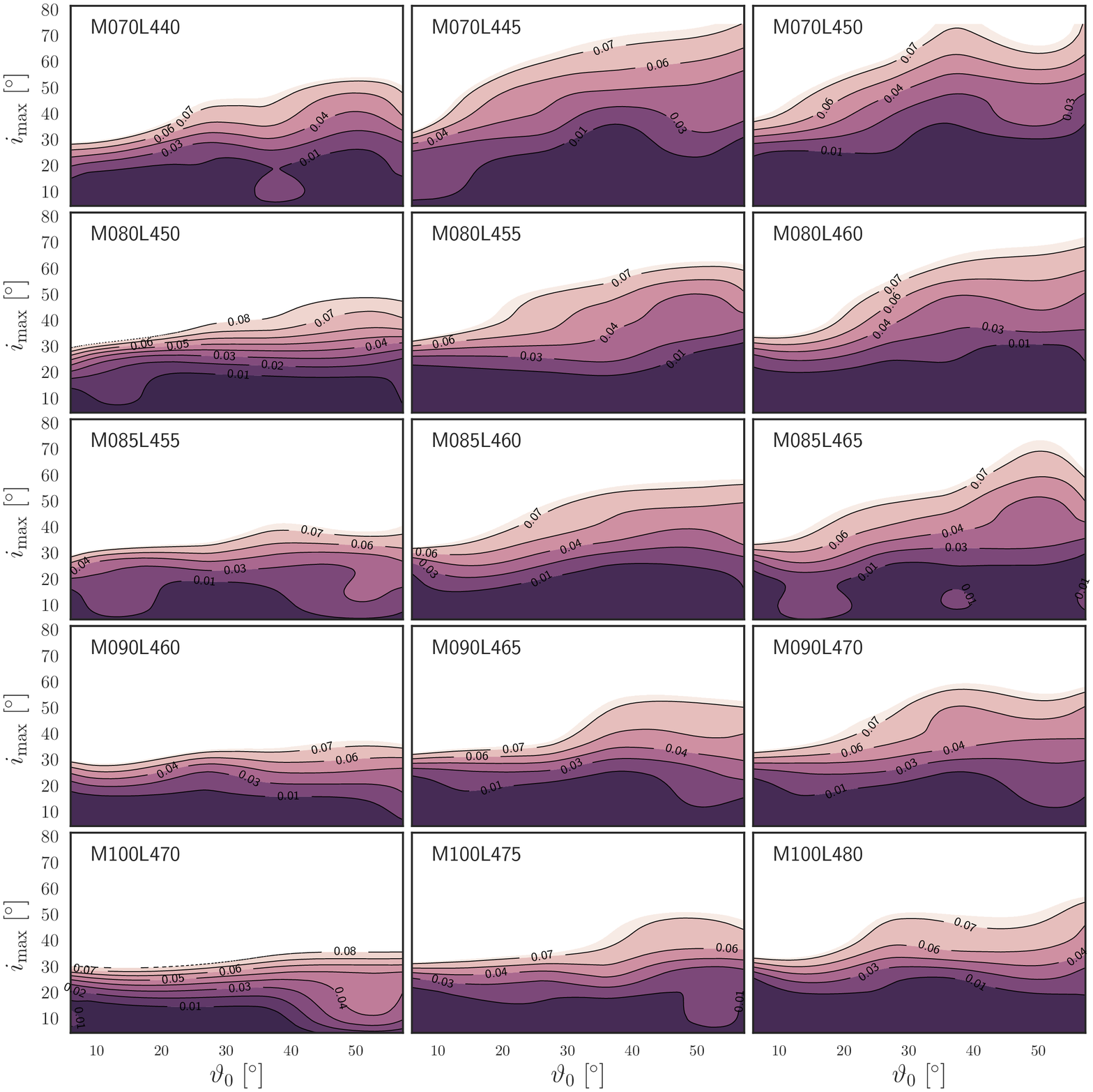} 
\caption{Contour plot of the standard deviation of the profile FWHM vs. launch 
and maximum inclination angles, for the $\lambda = 10$ case. 
Only the contours within the region matching the \citet{Fine+10} results are shown. 
Each panel corresponds to a different combination of mass and luminosity, labelled 
according to the adopted convention, as described in the caption to Figure 
\ref{Fig: LineProfiles-Lambda10-And-30-AccrRate01}. 
} 
\label{Fig: ContourPlot-i_maxVsTheta0-Lambda10}
\end{figure*}   

Recall that the region allowed by the \citet{Fine+10} result is a measure of the 
torus opening angle and the torus is, in the standard model, the component that 
determines whether an object seen under a viewing angle $i$ is Type  1 or 2. 
As can be seen from Figure \ref{Fig: ContourPlot-i_maxVsTheta0-Lambda10}, 
in the present framework our upper limit on the half-opening angle of the torus  
of an AGN depends on both the mass and luminosity of the central engine. 

If the broadening of the lines were only due to Keplerian motions modulated by the 
inclination dependence, the Fine et al. test would not differentiate among different 
mass-luminosity combinations, because that test analyses the dispersion of the 
logarithm of the line FWHMs rather than the FWHMs themselves. The fact that there 
are great diffe\-rences in the allowed regions in the $i_{\rm max}-\vartheta_0$ plane 
for the different cases is a consequence of a more complex entanglement between 
the relevant parameters (mass, luminosity, viewing and launch angles, acceleration 
mechanism, etc.).   

Looking along each row in Figure \ref{Fig: ContourPlot-i_maxVsTheta0-Lambda10} 
the mass is cons\-tant and the allowed regions in the $i_{\rm max}-\vartheta_0$ 
plane increase in size from left to right, as the Eddington ratio increases. 
We would like to link this to observational results other than those of \citet{Fine+10}.   
By construction, increasing the allowed region implies the possibility that the torus 
opening angle measured {\bf from the axis} also increases. In the scenario under 
consideration (fixed mass), that would imply that more luminous objects have a higher 
probability of being observed as Type 1 than less luminous counterparts. That is, 
there could be more torus opening angle values under which a luminous object would 
be classified as Type 1 than there are for fainter objects. 
Recent discussions in the literature suggest that this might the case. \citet{Elitzur+14} 
extended the analysis of \citet{Elitzur+Ho09} and confirmed the viability of a model 
where AGN broad-line emission follows an evolutionary sequence from Type 1 to 2 
as the accretion rate onto the central black hole is decrea\-sing. The authors suggest 
that the (at least partially) controlling parameter of this spectral evolution and the torus 
opening angle is $\Lbol/M^{2/3}$, that is only a function of $L$ for the 
$M = \text{const.}$ case.  

On the other hand, at fixed Eddington ratio (i.e., along a column in Figure 
\ref{Fig: ContourPlot-i_maxVsTheta0-Lambda10}), the allowed region decreases 
from top to bottom, i.e., with increasing mass. The \citet{Elitzur+14} results could 
be interpreted as implying that under fixed Eddington ratio conditions the torus 
opening angle (measured from the axis) should increase with decreasing mass.
That can be seen by rewriting the \citet{Elitzur+14} parameter in terms of the 
Eddington ratio as $\Lbol/M^{2/3} = \left(\Lbol/M\right) M^{1/3} 
\propto \left(\Lbol/\Ledd \right) M^{1/3} = \dot{m} \; M^{1/3}$. 
As mentioned, for the fixed Eddington ratio case, the results for our limits are opposite 
to \citet{Elitzur+14}.  
However, when considering fixed mass, our results match those of \citet{Elitzur+14}.  
An observational direct realization of such a case was discussed by \citet{LaMassa+15} 
as a plausible explanation for the ``changing look'' quasar they discovered. 

Note here that, in the case of \citet{Elitzur+14} framework, the decrement in observed 
broad lines originates in a decreasing accretion rate towards the central engine, that 
in turn decreases the outflow rate and the object's bolometric luminosity. 
At sufficiently low accretion rates, the BLR and the obscuring region are quenched, 
running out of fuel.  
 
We can also analyse how our findings can be explained in the context of a popular 
parametrization of the BLR size and the vertical size of the torus that is related to 
its half-opening angle, $\sigma_{\rm t}$. 
Some determinations of the vertical size of the torus \cite[e.g.,][]{Simpson05} provide 
$h_{\rm {t}} \propto L^{1/4}$.  
We can, then, obtain a crude estimate of the BLR opening using 
\citep[see e.g.,][]{Honig+Beckert07}
$\sigma_{\rm t} \sim  h_{\rm {t}}/R_{\rm BLR}  \sim L^{-1/4}$. 
Therefore,  more luminous objects would have smaller torus openings (measured from 
the disc). This is in agreement both with our results (at $M=$ const.) and those of 
\citet{Elitzur+14}. 

The dependence of the torus height on luminosity is a modification suggested by 
\citet{Simpson05} to the model known as ``receding torus'', proposed by \citet{Lawrence91}, 
wherein the height of the torus is constant as a function of radius. %
As discussed in the recent review by \citet{Bianchi+12}, this dependence of the 
obscuring structure covering factor on the luminosity has been supported by many 
observational results, in different wavelengths. For instance, hard X-ray studies 
\citep[e.g.,][]{Ueda+03, Akylas+06, Tozzi+06}
and in the optical  \citep[e.g.,][]{Arshakian05, Simpson05, Polletta+08}.  
\citet[][]{TU12, Alonso-Herrero+11} and more recently \citet{Oh+15} 
also show results consistent with the receding torus model.  
Other authors, on the other hand, do not favour it. \citet{LE10}, for instance, 
have questioned the validity of the above results, arguing that, at least in optical and 
IR-selected samples, such a luminosity dependence is an artifact due to the adopted 
definition of ``obscured'' and to the inclusion of low excitation AGN.

\subsection{Dispersion of $\log{\text{FWHM}}$-Clumpy torus}  
\label{sect: clumpy torus} 
In clumpy torus models, the obscuring structure is discrete, consisting of optically thick 
clouds and the quasar is obscu\-red when one such cloud is seen along the LoS. 
We start by briefly recalling the main characteristics of clumpy tori \citep[e.g.,][]{N+08a, Mor+09}. 
In such a model, the torus is characterised by the inner radius of the cloud distribution 
(set to the dust sublimation radius,  $R_{\rm  {d}}$, that depends on the grain properties 
and mixture) and six other parameters. These are the outer radius, $R_{\rm o}$; the 
viewing angle $i$; the torus width parameter (analogous to its opening angle), $\sigma$;  
the mean number of clouds along a radial equatorial line, $N_0$; the optical 
depth per cloud, $\tau_{\rm V}$ (the same for all clouds in the configuration) and the 
power-law index of radial density profile, $q$, such that the number of clouds fo\-llows 
$N(r) \propto r^{-q}$. 
Note that the outer radius is often given through the alternative parameter 
$Y = R_{\rm o}/R_{\rm d}$. 
An implicit assumption is that the disc and torus are aligned.  

Equation (3) in \citet{Mor+09} provides the escape probability associated with a soft-edge 
clumpy torus pres\-cription with a Gaussian distribution, that we include below for 
completeness  
\begin{equation} 
 P_{\rm  {esc}}(i) = \exp\left[{-N_{0}\exp\left({-\frac{(90\degr-i)^{2}}{\sigma^{2}}}\right)}\right]. 
 \label{eq: Pesc}
 \end{equation} 

Authors that support a clumpy rather than a smooth dust distribution torus 
\citep[e.g.,][]{Honig+Beckert07, N+08a} argue that the decreasing fraction of Type 2 
objects at high luminosities discussed in section \ref{sect: smooth torus} depends 
not only on the decreasing torus opening angle, but also on the decreasing $N_0$.  
The torus ``cove\-ring factor'' $f_2$ is estimated by the fraction of Type 2 sources in the 
total population \cite[e.g.,][]{LE10} and is related to the torus half-opening angle 
according to $f_2 = 1- \int_0^{\pi/2} P_{\rm esc}(i) \sin{i}~di$
For a smooth torus, this factor is given  by $f_2 =  \sin{\sigma_{\rm t}}$ but it is 
modi\-fied and becomes a function of the number of cloudets if a clumpy torus is 
considered. 

In Figure \ref{Fig: FWHMDisp-vs-Theta0-Nenkova+10-Probab} we show the standard 
deviation of the profile FWHM vs. launch angle obtained from our line profiles for 
the \citet{N+08a} clumpy torus formulation. Note that as we have adopted the soft-edge 
case of the model, only $i_{\rm  {max}}=90\degr$ is needed. Hence, the plots are curves 
and not contour levels. 
Each panel corresponds to a different combination of mass and luminosity and each line 
represents results for a different value of the torus model parameters $\sigma_{\rm t}$ 
and $N_0$ adopted from the \citet{Mor+09} sample. 
The black horizontal dashed line is the \citet{Fine+10} 0.08 dex constraint.  

As in Figu\-re \ref{Fig: ContourPlot-i_maxVsTheta0-Lambda10}, the mass is constant 
along rows and the accretion rate is constant along columns. 
As in the smooth torus case, we find that the agreement (or lack thereof) with the 
observational cons\-traint depends on the mass and the accre\-tion rate separately, 
with better agreement (i.e., reached from a smaller $\vartheta_0$ angle) achieved 
for the lowest mass and larger Eddington ratio. Indeed, at the lowest $\dot{m}$, 
there is no combination of parameters that satisfies the condition imposed.   
Additionally, the curves correspon\-ding to larger $N_0$ cross the limiting line at 
smaller $\vartheta_0$. That is, our results favour denser structures.   
   
\begin{figure*} 
\centering
\includegraphics[width=\textwidth]{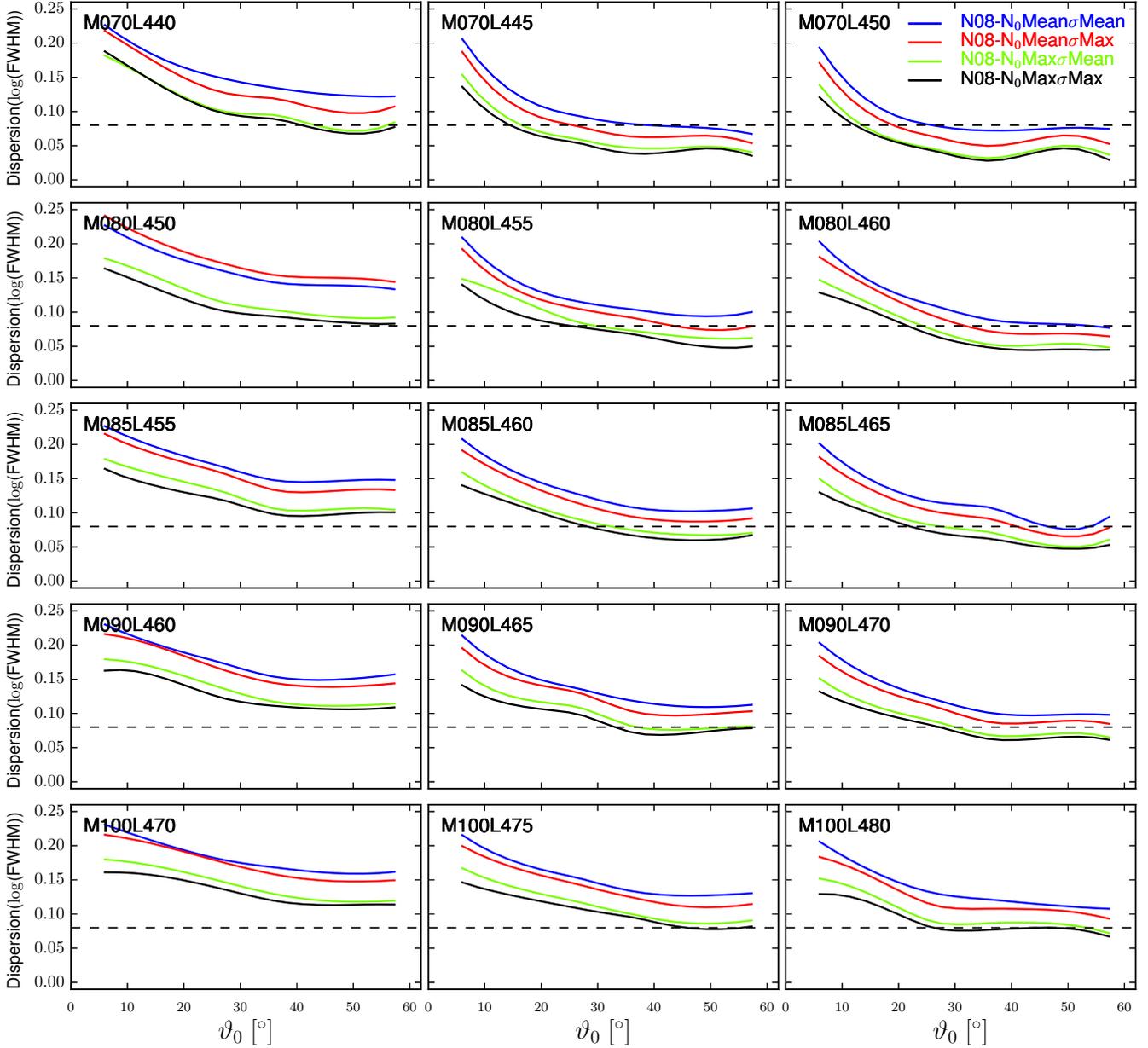} 
\caption{Standard deviation of the profile FWHM vs. launch angle, using the \citet{N+08a} 
clumpy torus prescription.  
The black horizontal dashed line corresponds to the \citet{Fine+10} 0.08 dex constraint. 
Each panel corresponds to a different combination of mass and luminosity and each line 
represent results for a given launch angle. 
} 
\label{Fig: FWHMDisp-vs-Theta0-Nenkova+10-Probab}
\end{figure*} 
 
\section{Comparing results for different $\lambda$ values} 
 \label{sect: lambda 30} 
In previous sections we have shown that the line profiles, and so their FWHMs and 
corresponding dispersions, depend only weakly on the dimensionless angular 
momentum parameter. 
Here we discuss this result in terms of the velo\-city and magnetic fields governing the 
line formation in the framework of our model. 
Recall that, within the ideal MHD formalism of wind launching sketched in Section 
\ref{subsect: self-sim sol.}, the angular momentum $l$ is constant along a field line 
(and so is its dimensionless version $\lambda$). 
Moreover, as also discussed in that Section, in the \citetalias{EBS92} model 
(and, therefore in ours), the adopted free parameters are $\lambda$ and $\vartheta_0$, 
while $\kappa$ is related to $\lambda$ by a nonli\-near correspondence, shown in 
Equation \eqref{eq: kappa_lambda}.
If we consider cases of equal masses and luminosities but diffe\-rent $\lambda$, 
and additionally assume same launch angles (i.e., same $\xi'_0$ values) and density 
structures, we can see that the difference is due only to differences in the magnetic fields. 
In effect, equal masses imply equal Keplerian velocities, which in turn implies that the 
differences in the para\-meter $k$ values can only be due to the magnetic field. 
Furthermore, equal masses also imply that the radial structure in the two cases are 
identical, so the difference in $\lambda$ is only related to the poloidal variable $\xi$. 

Comparing the profiles presented in Figures \ref{Fig: LineProfiles-Lambda10-And-30-AccrRate01} 
to \ref{Fig: LineProfiles-Lambda10-And-30-AccrRate1}, it is evident that the differences in 
shape and line-width bet\-ween results based on the two different $\lambda$ values adopted 
depend on $\vartheta_0$ but very mildly (if at all) on the viewing angle. 
For $i =  \text{const.}$, the difference is maximum for the minimum adopted value of 
$\vartheta_0$ and becomes negligible as the parameter increases.   
For $\vartheta_0 = \text{const.}$, the difference between profiles remain fairly constant 
along the whole range. There are a few departures, generally at the lowest $\vartheta_0$ 
and for the smallest $i$, but those do not invalidate the general behaviour. 

Similarly, the shapes and FWHMs show maximum differ\-ences for $\vartheta_0 =  6^{\circ}$ 
and almost negligible for larger $\vartheta$s, as can be seen from Figure  
\ref{Fig: Log10FWHMvsIncl}.   
The dispersion of linewidths, plotted in Figure \ref{Fig: VarvsInclAngle},  
show also that the larger differences \mbox{occur}, in general, at smaller $\vartheta_0$ values, 
with a few cases where noticeable differences are seen at large $\vartheta_0$ angles.  

The question is, then, why the results from the two diffe\-rent $\lambda$ values seem to be 
more distinguishable for smaller $\vartheta_0$? 
To analy\-se the issue, we will exploit the independence of the differences between profiles 
on viewing angle mentioned above. Note that, for given $M$ and $L$, these differences are 
also independent of $\nu$. The above fea\-tures allow us to choose convenient values for 
both frequency and $i$ that will simplify the analysis.  
We therefore adopt \mbox{$\nu = \nu_{\rm D}$ and $i = 0^{\circ}$} (although this particular 
viewing angle was not part of the simulations, the independence referred to allows this choice). 
Adopting $\nu = \nu_{\rm D}$ gives, replacing in Equations \eqref{eq: enu} and \eqref{eq: x2}, 
$e_\nu = e_{\nu_{\rm D}}=1$ and  \mbox{$x_\nu = x_{\nu_{\rm D}} = 0$}, respectively. Therefore, 
the optical depth, that is constructed using Equations \eqref{eq: tau} to \eqref{eq: x2} as  
$\tau_\nu = \tau~\!e_\nu~\!e^{-x_\nu^2}$,  
is simplified to $\tau_\nu \!= \!\tau_{\nu_{\rm D}} \!= \! \tau $.  
We also have $\vmath{D}(i= 0^{\circ}) = \vmath{p} \sin \vartheta$. 

To evaluate $\tau_{\nu_{\rm D}}$ we need to determine $Q$ and $q_{\rm tt}$ appropriate 
for the chosen frequency and viewing angle values. 
The expression of $Q$ is given in \citetalias{Ch+H13}. For $i = 0^{\circ}$, it simplifies to 
$Q = \Lambda_{zz} = \frac{\partial {\mathsf v{z}} }{\partial z}$. In 
the framework of the \citetalias{EBS92} solution to the MHD equations, this derivative is 
given by $ \sqrt{\frac{G M}{r_0^3}} \left[ \frac{f+4f'(\chi + c_2)}{2(\chi+2c_2)}\right]$. 
The first factor is independent of $\lambda$ and $\vartheta_0$, we can therefore concentrate 
only on the second factor. Recall that the profiles were obtained through quantities evaluated 
at $z =  z_{\rm em}$.   
Therefore, 
$Q(\chi = \chi_{\rm em}) \propto \left[ \frac{f(\chi_{\rm em})+
4f'(\chi_{\rm em})(\chi_{\rm em} + c_2)}{2(\chi_{\rm em}+2c_2)}\right]$. 
The leading factor of the dependence of the function $f(\chi)$ and its derivative on $\lambda$ is 
$f_\infty$, introduced in Equation \eqref{eq: f_infty}. 
This gives 
\begin{equation}
\label{eq: Q_approx} 
Q \sim \frac{f_\infty}{2 \left(\chi_{\rm em}+2c_2 \right)} = 
\sqrt{\frac{2 \lambda-3}{3}} \; \frac{1}{2 \left(\chi_{\rm em}+\tan \vartheta_0\right)} \,.
\end{equation}
Note that $q_{\rm tt}$ is intrinsicaly independent of $\lambda$. 
Thus, for the dependence on $\lambda$, $Q$ is the dominant term and we have, then, 
that the optical depth $\tau \sim Q^{-1}$ decreases when $\lambda$ increases. 
As $L_\nu \propto 1-e^{- \tau}$, we have then shown that  the luminosity would be larger 
for smaller $\lambda$. 
In addition, due to the term containing $\tan \vartheta_0$ in the numerator of Equation 
\eqref{eq: Q_approx}, the difference decreases with increasing $\vartheta_0$.  

\section{Discussion} 
\label{sect: discussion}
Considering the $\log{L}$-$\log{\mbox{FWHM}}$ plane in Figure \ref{Fig: simFig2-Fine10} 
and the contour plots of Disp($\log{\mbox{FWHM}}$) in 
Figure \ref{Fig: ContourPlot-i_maxVsTheta0-Lambda10} we would like to assess which 
parameter(s), and in which direction(s), should be changed in future simulations to achieve 
better agreement with the observational results.  

Note first that the angle $i_{\max}$ used throughout this work is measured from the 
polar axis (i.e, it is the complementary to the angle 
$\sigma_{\rm t} = \tan^{-1} h_{\rm t}/r_{\rm t}$ defined above). 
The receding torus model posits that the $i_{\max}$ parameter 
of more luminous objects is larger. 
Indeed, for constant mass our results permit that trend,  although they hint that the 
torus opening angle of more luminous objects is smaller than that of fainter counterparts.  
But, as mentioned in Section \ref{sect: smooth torus}, at fixed mass the trend is 
that reported by \citet{Elitzur+14}, i.e., that the torus ope\-ning angle decreases 
with luminosity. 
These points suggest that our model is, in its general conception, adequate although 
some of the parameter choices have to be reviewed for it to better represent real cases.  

For example, the tilt of the emitting region with respect to the equatorial plane was 
chosen to match that of \citetalias{MC97}, which was in turn chosen based on a BAL 
quasar fraction of $\sim 0.1$, a value that has been updated \citep{Allen+11}. 
More recent values put the BAL fraction as large as $\sim 0.25$ of the quasar 
population, implying an emission region with a steeper slope. In such a case, the 
effective length over which the emission is obtained would increase and be at 
higher distances above the disc. This would affect both the velo\-cities and their 
derivatives, thus affecting $Q$ and the optical depth, ultimately modifying the 
FWHM of the line profiles, but perhaps not as much their shapes.  
Here we recall that this slope can be a function of the radius, as opposed to the 
fixed value adopted in this work. 
Adopting a function that smoothly increases with radius (similar to the bowl-shaped BLR 
geometry in the model proposed by \cite{Goad+12}) instead of a constant slope, would 
make the streamlines intercept the emission region at higher heights from the disc plane, 
which would produce broader profiles, and probably with more dispersion. 
A related description of the BLR structure, described as ``nested rings''  
geometry, was proposed by \citet{Mannucci+92} and later reintroduced by Gaskell 
and collaborators \citep[e.g.,][]{Gaskell+08, Gaskell09}. 
Either of these geometries could provide a constraint to the families of possible 
curves to adopt as the base of the emission region. 

The spatial profile of the density could be reviewed or adjusted, too. Here we have to 
consider both the radial and the vertical structures, that are decoupled in the adopted 
model. 
As discussed in Section \ref{sect: Model Description}, the density structure in the radial 
direction is such that $n(r) \propto r^{-2}$, but a closer value to the corresponding 
\citet{SS73} rele\-vant accretion disc region would be a power-law exponent $-3/2$. 
Other possibilities involve changing the underlying disc model, adopting, e.g., a slim disc  
instead of the geome\-trically thin and optically thick disc of \citet{SS73}. 
Slim discs were developed in the pseudo-Newtonian limit by \citet{Abramowicz+88} 
and are more suitable for larger accretion rates ($\dot{m} \gtrsim 0.3$, e.g.,  
\citet{Abramowicz+Fragile13}). 

In the $z$ direction, the density drops off away from the base of the emission according 
to a half-Gaussian dependence, regulated by both the height and the thickness of the 
emission region. The former depends on the geometry of the region, discussed in the 
previous paragraph. Any modi\-fication to the latter should still ensure that it satisfies 
$l_{\rm em} \ll z_{\rm em}$. 
The radial size of the line-emitting region, given by the ratio $r_{\rm max}/r_{\rm min}$, 
was fixed throughout the present work. However, the amount of line shift and the degree 
of asymmetry of the line profiles are functions of it \citep[e.g.,][]{Flohic+12}. Therefore, it 
would be important to effectively study how this parameter affects the results. 

The turbulent velocity has been chosen to be constant, but it also can depend on the 
radial coordinate. 
For ins\-tance, it has been shown that magnetorotational instability \citep[MRI,][]{Balbus+Hawley91}  
provides a plausible mechanism to develop turbulence and transport angular momentum in 
discs, due to their differential Keple\-rian rotation. 

Another point to consider is that we did not explore changes in the photoionization 
calculations. The CLOUDY runs yielding the source function results used in this work 
were obtained adopting specific prescriptions for the various parameters in the code that 
define the photoionization state of the gas. The ionization and excitation structure of 
the gas depend on the spectral energy distribution (SED), the number density 
$n_\gamma$ of ionizing photons irradiating the gas, the metallicity and number density 
$n$ of the gas \citepalias[e.g.,][]{MC98}. 
The ionization parameter, defined as $U = n_\gamma/n$ encapsulates several of these 
quantities and is a convenient way to describe the ionization state of the medium. Thus, 
\begin{equation} 
U = \frac{1}{4 \pi r^2 c \;\! n_{\rm H}} 
\int_{\nu_0}^\infty \frac{L_\nu}{h \nu} d\nu = 
\frac{Q_{\rm H}}{4 \pi r^2  n \;\! c}, 
\end{equation}
where $\nu_0$ is the threshold ionization frequency, $h$ is 
Planck's constant, $n$ is the hydrogen number density and 
$Q_{\rm H} =  \int_{\nu_0}^{\infty} \frac{L_{\nu}}{h \nu}$ is the rate of ionizing photons 
emitted by the source. 
Values $U > 1$ indicate a highly ionized gas, while the gas is in low-ionization state 
for $U < 1$.   

Changes in the adopted prescription of any (or all) of these parameters will lead to a 
different structure of the gas and its ionization and excitation states, in the disc and 
wind. 
For example, for most of their model cases \citet{MC97, MC98} employed a (modi\-fied 
in the X-ray region) \citet{Mathews+Ferland87} SED, that is still very popular in the 
literature.  
Adopting a different SED would affect the ionization structure in the wind, but in a wind 
scenario there is not total freedom in choosing an alternative. A distribution too strong 
in X-rays compared with the UV portion may preclude the formation of a wind, 
because the gas is overionized before it can be accelerated. In such a case, the higher 
ionization lines, such as \ion{C}{iv}, would instead be produced in the low-velocity gas, 
that is illuminated by a continuum that is now strong in the extreme UV and in the X-rays  
\citep{Leighly04}. 

In section \ref{subsect: self-sim sol.}  we mentioned that our 
work does not account for scattering effects. However, the importance of this process 
has been recently highlighted by \mbox{\citet{Higginbottom+14}} who, using 
Monte-Carlo radiative transfer simulations, showed that including scattering of 
ionizing photons leads to a very high ionization parameter and modifies the outflow 
emission properties. 

\section{Summary and Conclusions} 
\label{sect: Conclusions}

In this work we have studied AGN broad emission line profiles with a model 
that combines an improved version of the accretion disc wind model of \citet{MC97} 
with the hydromagnetic driving of \citet{EBS92}.  
The dynamics of these self-similar MHD outflows is characterised by two 
parameters, e.g., the dimensionless angular momentum $\lambda$, and 
the wind-launch angle with respect to the disc plane $\vartheta_0$.  

We have compared the dispersions in our model \ion{C}{iv}~linewidth distributions to 
observational upper limits on that dispersion. Those limits translate to an upper 
limit to the half-opening angle of the putative torus feature that is part of the 
standard model describing the AGN phenomenon. 
To achieve this, we constructed contour plots of the dispersion of $\log(\mbox{FWHM})$  
in the $\vartheta_0-i_{\rm max}$ plane, capped with the \citet{Fine+10} observed 
upper limit dispersion, defining in this way a boundary line 
$i_{\rm max}$ vs $\vartheta_0$, below which an object can be seen as Type 1. 

The maximum torus half-opening angle of about 47$\degr$ reported in \citetalias{Ch+H13}  
has been corrected.  That value was obtained based on an erroneous interpolation 
routine of the CLOUDY-generated source function points. 
The revised maximum torus half-opening angle is larger, about 75$\degr$. 
In fact, the maximum torus half-opening angle is an increasing function of the 
wind launch angle $\vartheta_0$.

We extended the analysis presented in \citetalias{Ch+H13} to consider a range of 
black hole masses and luminosities. In a similar manner to the approach adopted 
in the fiducial case, we computed, for different combinations of mass and luminosity 
of the central object within that range, line profiles corresponding to the same 
combinations of wind-launch and viewing angles used before. 
Additional series of model runs with different values of $\lambda$ suggest that the 
profile linewidths and corresponding dispersions depend only mildly on this parameter. 

We also found that many of the profile line characte\-ristics, such as the FWHM, 
the (blue- or red-) shift with res\-pect to the systemic velocity, and the degree of 
asymmetry, depend not only on the viewing angle (a parameter external to the object, 
depen\-ding on the orientation of the observer relative to the source), 
but also on the launch angle $\vartheta_0$, a parameter that is intrinsic to the object. 
Additionally, our results suggest that large values of $\vartheta_0$ are preferred. 

The analysis of the luminosity-linewidth relation showed that the black hole masses 
could not be recovered if the relation proposed by \citetalias{VP06} (reproduced in 
Equation \eqref{eq: VP06-eq7}) is applied. 
While this discrepancy could be due to the FWHM of our line profiles, it would be worth 
evaluating if the inclusion of new empirical data from the literature supports the view 
that the expression needs to be reconsidered.

We then studied, for each case, the dispersion of $\log(\mbox{FWHM})$ and 
imposed the observational results of \citet{Fine+10}. 
Again, contour plots of the dispersion of $\log(\mbox{FWHM})$ in the 
$\vartheta_0-i_{\rm max}$ plane, constrained with the \citet{Fine+10} results, 
were used to determine the torus half-opening angle appropriate for each case.  
The picture that emerged is that the dispersion of the linewidth depends on both 
the mass of the central object and the Eddington ratio at which it is fed. 
At fixed mass, the maximum allowed torus half-opening angle (measured from the 
axis) increases with increasing Eddington ratio (equivalent, under the fixed mass 
condition, to increasing luminosity). 
That can be interpreted as objects with larger accretion rates having a higher 
probability of being observed as Type 1 AGN than those fed at a lower rate. 
At fixed Eddington ratio, the probability that an object will be seen as Type 1 
decreases with increasing mass.  
But the timescales for mass changes due to accretion are larger enough for any 
particular system to be considered of constant mass, and, as mention above, in 
that scenario our results are in agreement with the observational report of \citet{Elitzur+14}. 

A clumpy torus was also analysed. The limiting viewing angle to integrate over this 
was always 90$^{\circ}$, as opposed to the differential $i_{\rm max}$ that was used 
in the putative smooth torus case. 
Except for this difference, this alternative obscuring structure yielded similar 
constraints to those obtained with the traditional torus.  
Moreover, the results favour denser structures.   

Ultimately, this work links theory with observational results, by imposing an observational 
constraint to the distribution of a property of the emission lines emerging from the BLR. 
The BLR is modelled through a physically well motivated wind description and the 
observational constraint on the dispersion of the line-width distribution allows it to be 
translated into a constraint on the geometry of the obscu\-ring structure that is invoked 
to explain the Type 1/Type 2 dichotomy among AGNs. 

\section*{Acknowledgments}
LSC and PBH acknowledge support from NSERC. 
The authors would like to thank the anonymous referee for 
very helpful comments and suggestions.

\bibliographystyle{mnras}
\bibliography{ref-Chajet-2016}


\bsp	
\label{lastpage}
\end{document}